\documentclass[aps,pre,reprint,showpacs,superscriptaddress,eqsecnum]{revtex4-1}

\usepackage{combelow}
\usepackage{bm} % \bm command for vectors
\usepackage{braket} % \braket for < > notation
\usepackage{amsmath} % \eqref command
\usepackage{amsfonts}
\usepackage{amssymb}
\usepackage{graphicx}
\usepackage{epstopdf}

\usepackage[normalem]{ulem} % REMOVE BEFORE submission to PRE

\usepackage{enumitem}

\usepackage{xcolor}

\def\feq{\ensuremath{f^{(\mathrm{eq})}}}

\def\wp{{\overline{p}}}
\def\vp{{\bm{p}}}

\def\vp{{\bm{p}}}
\def\vu{{\bm{u}}}
\def\vxi{{\bm{\xi}}}
\def\vq{{\bm{q}}}

\begin{document}

\title{Half-range lattice Boltzmann models for the simulation of 
Couette flow using the Shakhov collision term}

\author{Victor E. \surname{Ambru\cb{s}}}
\email[E-mail: ]{victor.ambrus@e-uvt.ro}
\affiliation{Center for Fundamental and Advanced Technical Research,
Romanian Academy, Bd.~Mihai Viteazul  24, 300223 Timi\cb{s}oara, Romania}
\affiliation{Department of Physics, West University of Timi\cb{s}oara,\\
Bd.~Vasile P\^arvan 4, 300223 Timi\cb{s}oara, Romania}

\author{Victor \surname{Sofonea}}
\email[E-mail: ]{sofonea@gmail.com, sofonea@acad-tim.tm.edu.ro}
\affiliation{Center for Fundamental and Advanced Technical Research,
Romanian Academy, Bd.~Mihai Viteazul  24, 300223 Timi\cb{s}oara, Romania}

\date{\today}

\begin{abstract}
The three-dimensional Couette flow between parallel plates is addressed using mixed lattice Boltzmann 
models which implement the half-range and the full-range Gauss-Hermite quadratures on the Cartesian axes 
perpendicular and parallel to the walls, respectively. The ability of our models to simulate 
rarefied flows are validated through comparison against previously reported results obtained 
using the linearized Boltzmann-BGK equation for values of the Knudsen number (Kn) up to $100$. 
We find that recovering the non-linear part of the velocity profile (i.e., its deviation 
from a linear function) at ${\rm Kn} \gtrsim 1$ requires high quadrature orders.
We then employ the Shakhov model for the collision term to obtain 
macroscopic profiles for Maxwell molecules using the standard 
$\mu \sim T^\omega$ law, as well as for
monatomic Helium and Argon gases,
modeled through ab-initio potentials,
where the viscosity is recovered using the Sutherland model.
We validate our implementation by comparison with DSMC results 
and find excellent match for all 
macroscopic quantities for ${\rm Kn} \lesssim 0.1$. At ${\rm Kn} \gtrsim 0.1$, 
small deviations can be seen in the profiles of the diagonal components of the pressure tensor,
the heat flux parallel to the plates, and the velocity profile, as well as in
the values of the velocity gradient at the channel center.
We attribute these deviations to the 
limited applicability of the Shakhov collision model for highly out of equilibrium 
flows.
\end{abstract}

\maketitle

\section{Introduction}

It is generally recognized that the Navier-Stokes-Fourier
equations are not appropriate to investigate the flow phenomena in
highly rarefied gases where the continuum hypothesis is no longer valid.
To investigate such far from equilibrium fluids, the Boltzman
equation, which governs the evolution of the one-particle distribution function 
$f \equiv f(\bm{x}, \bm{p}, t)$ in a seven-dimensional space, can be employed
instead \cite{grad58,kogan69,harris71,
gatignol75,cercignani88,bird94,sone02,liboff03,karniadakis05,shen05,sone07,tabeling11,sharipov16}.
 Finding solutions of the Boltzmann equation is a challenging task due to the 
complexity of the collision term, which requires the evaluation of $5$-dimensional 
integrals over the momentum space. Effective approaches to solve the Boltzmann equation 
include the celebrated direct simulation Monte Carlo (DSMC) method \cite{bird94,sharipov16}, where 
the collision integral is sampled by considering a sufficiently large ensemble of 
representative particles which are evolved individually; the 
discrete velocity method (DVM) \cite{broadwell64,sone02,aoki03,sone07,sharipov16}; 
and more recently, the fast spectral 
method, which relies on the projection of the collision term on orthogonal functions
\cite{mouhot06,filbet12,wu13,gamba17}. In all 
approaches mentioned above, the evaluation of 
the collision term still remains 
the most time-consuming part of the numerical algorithm, placing severe constraints on the 
size and complexity of the systems which can be analyzed numerically.

In the early '50s, Bhatnaghar, Gross and Krook introduced their
single relaxation time approximation of the collision term
of the Boltzmann equation describing the ideal gas.
This approximation, known as the BGK model, was 
derived under the assumption that the deviation of the 
gas from the local (Maxwellian) equilibrium is small \cite{bhatnagar54}. The severe 
drawback of the BGK model is that the transport coefficients are 
governed by the single
relaxation time $\tau$, and in particular, the 
Prandtl number ${\rm Pr}$ is fixed at $1$. This limitation was overcome 
through the collision term model proposed by Shakhov 
\cite{shakhov68a,shakhov68b,yang95,li05}, who extended the
single relaxation time (BGK) model
to allow ${\rm Pr}$ to be controlled 
independently from the relaxation time $\tau$.
Other extensions of the BGK model,
 which allow ${\rm Pr}$ to be controlled, include 
the ellipsoidal BGK (E-BGK) model \cite{holway66,meng13,zhang18} and the multi-relaxation time 
(MRT) models widely employed in lattice Boltzmann simulations \cite{dhumieres92,dhumieres02}.
The Shakhov model (also known as the S-model) 
was subsequently extended by Rykov et al. to account for 
rotational degrees of freedom \cite{rykov75}.
In the linearized regime, the Gross-Jackson \cite{gross59} and the 
McCormack \cite{mccormack73} models extended
 the relaxation time paradigm to account for
realistic interaction cross-sections for single gases and 
for gaseous mixtures, respectively.

The simplicity of the relaxation time formulation 
motivated researchers to implement and develop such models and a surprising 
range of effects turned out to be correctly recovered. It is now generally 
accepted that the relaxation time approach can be used to simulate flows 
which are not far from equilibrium, provided the transport coefficients 
are correctly recovered \cite{ambrus12,sharipov16,titarev18}. 
Solutions of these so-called Boltzmann model equations
can be obtained using a variety of numerical
methods, amongst which we recall 
the DVM \cite{ho15,sharipov16}, the discrete unified gas-kinetic scheme (DUGKS)
\cite{guo13pre,guo15pre,peng18caf,zhu17cpc,zhu16caf},
the lattice Boltzmann (LB) models
\cite{aidun,chen98,sukop06,shan06,succi01,gladrow00,kruger17,succi18},
the off-lattice Boltzmann models (OLBM) \cite{nannelli92,lee03,cheng04,min11,fakhari15,kramer17}, 
and the discrete Boltzmann models (DBM) \cite{he98pre,li03prl,xu12,xu16,lin17,xu18}.

In this paper, we reasses the capabilities of the Shakhov collision model 
in the context of the three-dimensional ($3D$)
Couette flow between parallel plates by employing
lattice Boltzmann (LB) models based on the Gauss quadrature method.
From a historical perspective, the LB models emerged as successors of the 
lattice gas automata \cite{succi01,gladrow00} introduced more than three 
decades ago and were originally designed to recover the Navier-Stokes equations
using a fast and simple algorithm. The efficiency of
this algorithm relies on the implementation of the advection and time stepping
according to the \emph{collide-and-stream} paradigm
involving the discretization of the momentum space using a
relatively small number of
momentum vectors which exactly connect neighboring lattice 
sites \cite{aidun,chen98,he97a,he97b,succi01}. 
Besides their numerical efficiency,
the key to the success of these early LB models
 is the implementation of the 
collision term using a polynomial series which allows its moments to 
be exactly recovered up to a certain order $N$ \cite{shan06}. 
While astonishingly successful at the Navier-Stokes level 
\cite{aidun,chen98,sukop06,shan06,succi01,gladrow00,kruger17,succi18}, 
the LB models based on the collide-and-stream paradigm obtained only 
limited success when applied to flows of rarefied gases \cite{shi11,meng11jcp,shi15}. 
In particular, it was noted that in order to achieve accurate results 
at non-negligible values of the Knudsen number ${\rm Kn}$, the 
velocity set must be enriched to account for higher order
moments of the distribution function. This brings about 
a series of complications, since during a single time step,
the particles must hop over an increasing number of lattice sites 
\cite{chikatamarla09,ansumali16,philippi06,sbragaglia07}.
This also renders the implementation of boundary conditions cumbersome 
\cite{meng11jcp,meng14}.

A straightforward alternative to the collide-and-stream paradigm comes from 
the DVM implementations of the Boltzmann equation, where the advection and 
time stepping are implemented using finite differences. Retaining the 
simplified polynomial truncation of the collision term, these implementations
can be referred to as finite difference lattice Boltzmann (FDLB) models 
\cite{cao97pre,chen98pre,he97ijmpc,mei98jcph,ubertini03,watari04,bardow08,gan11,meng11pre,meng11pre2}, or 
discrete Boltzmann models \cite{he98pre,li03prl,xu12,xu16,lin17,xu18}. 
Amongst the first high-order FDLB 
models are the 2D shell-based models introduced by Watari and Tsutahara 
\cite{watari03,watari07,watari10,watari16}
and their 3D generalizations \cite{watari06,watari09,romatschke11,ambrus12},
the models based on the tensor Hermite polynomials
\cite{shan06,meng11pre,meng11pre2},
and, more recently, the models based on the Cartesian split of the velocity space 
discussed in
Refs.~\cite{ambrus14ijmpc,ambrus14ipht,ambrus14pre,ambrus16jcp,
ambrus16jocs}.

The Couette flow between parallel plates has become a benchmark problem
for fluid dynamics simulations. In the context of rarefied 
gases, the linearized Boltzmann-BGK equation has been solved 
in a semi-analytic manner to high numerical precision in 
Ref.~\cite{jiang16} (see also Refs.~\cite{sone07,yap12,li15} for previous 
results). The LB results obtained within the collide-and-stream paradigm 
reported excellent agreement with DVM or DSMC results for small values of 
${\rm Kn}$, but their performance quickly deteriorated as ${\rm Kn}$ was 
increased towards the transition regime \cite{ansumali07,yudistiawan08,
yudistiawan10,shi11,meng11jcp,shi15}. There has been significant effort
devoted to 
deriving macroscopic equations which account for the non-equilibrium 
features appearing at non-negligible ${\rm Kn}$, from which we only mention 
the regularized 13 moments system of equations (R13) 
\cite{struchtrup07,torrilhon08,taheri09,struchtrup11,rana16}.
The limited success of the R13
system in the transition regime suggests that even more moments should be 
taken into account. From a lattice Boltzmann perspective, this 
is equivalent to extending the quadrature order of the model. 
Indeed, this was confirmed in Ref.~\cite{ambrus12}, when it was shown that 
FDLB models based on the spherical factorization of the momentum space 
exhibit a slow but steady convergence with respect to the quadrature 
order at ${\rm Kn} = 0.5$.

Recent FDLB studies of the Couette flow showed that accurate solutions of 
the S-model equation 
can be obtained by enriching the velocity set 
\cite{ambrus12,ambrus14ijmpc,ambrus14pre,ambrus16jcp}. 
The crucial piece which makes the simulations 
much more efficient is to take into account the discontinuity in the 
distribution function, which is induced by the boundary conditions 
prescribed at the solid walls \cite{gross57,takata13}. For this reason,
appropriate half-range quadratures should be employed on the direction perpendicular to the walls
\cite{yang95,li03,li04,li09,lorenzani07,frezzotti09,frezzotti11,
gibelli12,guo13pre,ghiroldi14,ambrus14pre,ambrus14ipht,ambrus14ijmpc,
guo15pre,gibelli15,shi15,ambrus16jcp,ambrus16jocs}.
In this paper, we will employ the mixed quadrature LB models introduced in 
Ref.~\cite{ambrus16jcp} for the study of the 2D Couette flow.

Since we are interested in obtaining accurate results for the temperature field, 
as well as for the heat flux, the number of degrees of freedom in the 
momentum space must be equal to three.
In the context of flows which are effectively two-dimensional
(e.g., in the $xy$ plane), 
this can be achieved by working with reduced distributions
\cite{li04,meng13jcp}. These distributions are obtained by 
analytically integrating the momentum space degree of freedom along the 
direction which 
is perpendicular to the normal to the walls and to the direction of the flow.
The ensuing LB models employ the half-range 
Gauss-Hermite quadrature on the axis perpendicular to the walls 
(the $x$ axis) and the full-range Gauss-Hermite 
quadrature for the axis parallel to the wall (the $y$ axis).
The advantage of these
 LB models with {\emph{mixed quadratures}}
 is described as follows. 
The discontinuity induced by the diffuse reflection boundary conditions imposed on the walls 
perpendicular to the $x$ axis warrants the use of the half-range 
Gauss-Hermite quadrature, which requires  $Q = N + 1$ points on each
Cartesian semiaxis to ensure the accurate recovery of the moments of 
the distribution function $f$ up to $N$th order.
The $2Q$ quadrature points on the whole axis are
twice the number of points required by the full-range Gauss-Hermite
quadrature to achieve the same degree of accuracy when considering 
the full-space moments of $f$ \cite{ambrus16jcp}.
Thus, it is more convenient to employ
the full-range Gauss-Hermite quadratures 
on the $y$ direction (i.e., the direction parallel to the walls), where
no discontinuities in the  distribution function arise,
resulting in an overall smaller velocity set.

The aim of this paper is to demonstrate that the Shakhov
collision term model can be 
successfully employed in LB simulations to 
match Direct Simulation Monte Carlo (DSMC) results. 
In particular, we consider comparisons with the data 
obtained for the Couette flow of Maxwell molecules 
\cite{struchtrup07,struchtrup08,torrilhon08,taheri09,schuetze03},
as well as for the flow of Helium and Argon modeled using ab initio 
potentials \cite{sharipov13}.

For the spatial advection, we employed the fifth order weighted essentially 
non-oscillatory (WENO-5) 
scheme described in 
Refs.~\cite{jiang96,gan11,blaga17prc,hejranfar17pre,busuioc17}, while the 
time stepping was performed using a third-order Runge-Kutta algorithm 
\cite{shu88,gottlieb98,henrick05,trangenstein07,blaga17prc,busuioc17}. The 
implementation of the diffuse 
reflection boundary conditions on the channel walls is identical to that 
presented in Ref.~\cite{busuioc17}.

The outline of this paper is as follows. 
In Sec.~\ref{sec:shk}, the Shakhov collision term is 
briefly described.
Section~\ref{sec:lb} is dedicated to discussing the Boltzmann equation 
for the $3D$ Couette flow between 
parallel plates, in the context of reduced distributions. 
Section~\ref{sec:lowmach} validates our models in the 
incompressible (low Mach) regime, by comparison with the semi-analytic 
benchmark data reported in Ref.~\cite{jiang16}.
In Sec.~\ref{sec:mw}, our models are validated against the 
DSMC results for Maxwell molecules reported in
Refs.~\cite{struchtrup07,struchtrup08,torrilhon08,taheri09,schuetze03}, 
at various values of ${\rm Kn}$ and of the wall velocity $u_w$.
In Sec.~\ref{sec:sharipov}, we consider a comparison with the 
DSMC simulation results based on ab-initio potentials for Helium and Argon
\cite{sharipov13}.
Section~\ref{sec:conc} concludes this paper.
The details regarding our numerical scheme are presented in Appendix~\ref{app:scheme}.
We warn our readers that, in order to facilitate the comparison of our LB results with 
various data in the literature, the notation used to refer to the degree 
of rarefaction varies between the sections where these results are reported,
as follows: in Sec.~\ref{sec:lowmach}, the notation $k$ is employed 
to refer to the Knudsen number; in Sec.~\ref{sec:mw}, the Knudsen number
 is denoted 
using the familiar notation ${\rm Kn}$ and is linked to $k$ through
 ${\rm Kn} = k / \sqrt{2}$;
in Sec.~\ref{sec:sharipov}, the degree of rarefaction is characterized using the 
rarefaction parameter $\delta$, which is linked to $k$ (from Sec.~\ref{sec:lowmach}) 
and ${\rm Kn}$ (from Sec.~\ref{sec:mw}) through $\delta = 1/k$ and $\delta = 
1/ {\rm Kn} \sqrt{2}$, respectively.

\section{Boltzmann equation with the Shakhov collision term}\label{sec:shk}

In this section, we establish our notation by introducing the Shakhov collision 
term model, as well asour non-dimensionalization convention.

\subsection{Shakhov model}\label{sec:shk:model}

The Boltzmann equation for a force-free flow with the Shakhov
collision term is given by
\cite{shakhov68a,shakhov68b,guo15pre,titarev07,graur09,ambrus12}:
\begin{equation}
 \partial_{\widetilde{t}} \widetilde{f} + 
 \frac{\widetilde{\vp}}{\widetilde{m}} \cdot \widetilde{\nabla} 
 \widetilde{f}
 = - \frac{1}{\widetilde{\tau}}\left[\widetilde{f} - 
 \widetilde{f}^{({\rm eq})} \left(1+{\mathbb{S}}\right)\right],
 \label{eq:boltz_dim}
\end{equation}
where the overhead tilde $\widetilde{\phantom f}$ denotes dimensionful quantities.
In the above, $\widetilde{f}$ is the Boltzmann distribution function, 
$\widetilde{m}$ and $\widetilde{\vp}$ are the particle mass and momentum vector, respectively,
while $\widetilde{\tau}$ is the relaxation time. 
The Maxwell-Boltzmann equilibrium distribution function is 
given by:
\begin{equation}
 \widetilde{f}^{({\rm eq})} = \frac{\widetilde{n} }{(2\pi \widetilde{m} \widetilde{k}_B \widetilde{T})^{3/2}} 
 \exp\left[-\frac{(\widetilde{\vp} - \widetilde{m}\widetilde{\vu})^2}{2\widetilde{m} \widetilde{k}_B \widetilde{T}}\right],
 \label{eq:feq_dim}
\end{equation}
where $\widetilde{n}$, $\widetilde{T}$ and $\widetilde{u}$ are the macroscopic
density, temperature and velocity of the fluid. 
The Shakhov term $\mathbb{S}$ is given by:
\begin{equation}
\mathbb{S} = \frac{1-{\rm Pr}}{\widetilde{n} \widetilde{K}_B^2 \widetilde{T}^2}
\left(\frac{\widetilde{\vxi}^{2}}{5\widetilde{m}\widetilde{K}_B \widetilde{T}}-1\right)
\widetilde{\vq} \cdot \widetilde{\vxi},\label{eq:sdef}
\end{equation}
where ${\rm Pr}$ gives the Prandtl number (see below), 
$\widetilde{\vxi} = \widetilde{\vp} - \widetilde{m}\widetilde{\vu}$ 
is the peculiar momentum, while $\widetilde{\vq}$ is the heat flux.

The macroscopic quantities $\widetilde{n}$, $\widetilde{\vu}$, 
pressure tensor $\widetilde{T}_{ij}$ and $\widetilde{\vq}$ can be obtained 
as moments of $\widetilde{f}$:
\begin{align}
 \widetilde{n} =& \int d^3\widetilde{p} \, \widetilde{f},\nonumber\\
 \widetilde{\vu} =& \frac{1}{\widetilde{\rho}} \int d^3\widetilde{p} \, \widetilde{f}\, \widetilde{\vp},
 \nonumber\\ 
 \widetilde{T}_{ij} =& \int d^3\widetilde{p} \, 
 \widetilde{f}\, \frac{\widetilde{\xi}_i \widetilde{\xi}_j}{\widetilde{m}},\nonumber\\
 \widetilde{\vq} =& \int d^3\widetilde{p}\, \widetilde{f} \frac{\widetilde{\vxi}^2}{2\widetilde{m}} 
 \frac{\widetilde{\vxi}}{\widetilde{m}},\label{eq:qdef}
\end{align}
while the pressure is obtained as $\widetilde{P} = \frac{1}{3}(\widetilde{T}_{xx} +
\widetilde{T}_{yy}+\widetilde{T}_{zz}) = \widetilde{n} \widetilde{K}_B \widetilde{T}$.

By employing the Chapman-Enskog expansion, it can be seen that, for three-dimensional flows, 
the Shakhov collision term gives rise to the following expressions for the transport coefficients,
namely the dynamic (shear) viscosity $\widetilde{\mu}$ and the heat conductivity 
$\widetilde{\kappa}_{T}$ \cite{ambrus12}:
\begin{equation}
 \widetilde{\mu} = \widetilde{\tau} \widetilde{n} \widetilde{K}_B \widetilde{T}, \qquad 
 \widetilde{\kappa}_{T} = \frac{1}{{\mathrm{Pr}}} \frac{5\widetilde{K}_B}{2\widetilde{m}}
 \widetilde{\tau} \widetilde{n} \widetilde{K}_B \widetilde{T},
\end{equation}
where the Prandtl number ${\rm Pr}$, calculated as
\begin{equation}
{\rm Pr} = \frac{\widetilde{c}_{p}\widetilde{\mu}}{\widetilde{\kappa}_{T}}, 
\qquad \widetilde{c}_{p} = \frac{5\widetilde{K}_B}{2\widetilde{m}},
\end{equation}
is adjustable according to Eq.~\eqref{eq:sdef}.

\subsection{Non-dimensionalization convention}\label{sec:shk:nondim}

Our non-dimensionalization convention follows the one employed in Ref.~\cite{sofonea05},
being based on the following reference quantities: the reference length 
$\widetilde{l}_{\rm ref} = \widetilde{L}$, the reference temperature
$\widetilde{T}_{\rm ref}$, the reference mass $\widetilde{m}_{\rm ref}$, and
the reference density $\widetilde{n}_{\rm ref}$.
In the context of the Couette flow, the reference length is taken as the 
distance between the parallel plates, $\widetilde{l}_{\rm ref} = \widetilde{L}$,
and the reference temperature is the wall temperature, 
$\widetilde{T}_{\rm ref} = \widetilde{T}_w$.
The reference mass is taken equal to the particle mass $\widetilde{m}_{\rm ref}$,
while the reference speed is:
\begin{equation}
 \widetilde{c}_{\rm ref} = \sqrt{\frac{\widetilde{K}_B \widetilde{T}_{\rm ref}}
 {\widetilde{m}_{\rm ref}}} = \frac{1}{\sqrt{\gamma}} \widetilde{c}_{s;{\rm ref}},
 \label{eq:cref}
\end{equation}
where $\widetilde{c}_{s;{\rm ref}}$ is the sound speed at the reference temperature 
and $\gamma$ is the adiabatic index.
Since we only consider monatomic ideal gases ($\gamma = 5/3$), 
the Mach number corresponding to the non-dimensionalized
velocity $u$ is:
\begin{equation}
 {\rm Ma} = u \sqrt{\frac{3}{5}} \simeq 0.775 u.
 \label{eq:Ma}
\end{equation}

The non-dimensionalized distribution function $f$ is defined as:
\begin{equation}
 f = \frac{\widetilde{f}}{\widetilde{n}_{\rm ref}} 
 (\widetilde{m}_{\rm ref} \widetilde{k}_B \widetilde{T}_{\rm ref})^{3/2},
\end{equation}
such that the non-dimensional form of the
Maxwell-Boltzmann distribution \eqref{eq:feq_dim}
is:
\begin{equation}
 \feq = \frac{n}{(2\pi mT)^{3/2}} 
 \exp\left[-\frac{(\vp - m\vu)^2}{2mT}\right],
 \label{eq:feq}
\end{equation}
where $\vp \equiv \widetilde{\vp} / \widetilde{m}_{\rm ref} \widetilde{c}_{\rm ref}$ 
and the non-dimensional mass $m = \widetilde{m} / \widetilde{m}_{\rm ref} = 1$ is kept 
explicitly for the sake of clarity of the mathematical relations presented in 
what follows.
Finally, the Boltzmann equation \eqref{eq:boltz_dim} can be non-dimensionalized after 
multiplying both sides by $\widetilde{t}_{\rm ref} 
(\widetilde{m}_{\rm ref} \widetilde{k}_B \widetilde{T}_{\rm ref})^{3/2} / \widetilde{n}_{\rm ref}$,
yielding:
\begin{equation}
 \partial_t f + \frac{\vp}{m} \cdot \nabla f = -\frac{1}{\tau}
 \left[f - \feq\left(1+\mathbb{S}\right)\right],
 \label{eq:boltz}
\end{equation}
The exact expression for the nondimensionalized relaxation time $\tau$ 
depends on the collision term
model and will be discussed separately in Secs.~\ref{sec:lowmach}, 
\ref{sec:mw} and \ref{sec:sharipov}.

\section{Lattice Boltzmann models for the Couette flow}\label{sec:lb}

In the remainder of this paper, we focus on the Couette flow between parallel plates.
The coordinate system is chosen such that the $x$ axis is perpendicular to the plates, which are 
located at $x_{\rm left} = -L / 2$ and $x_{\rm right} = L / 2$, where the non-dimensionalized 
channel length $L = 1$ is kept explicitly in order to facilitate the physical interpretation 
of the mathematical expressions appearing below. The plates are set in motion 
along the $y$ direction with constant velocites $u_{\rm left} = -u_w$ and $u_{\rm right} = u_w$, 
while their temperature is kept constant ($T_{\rm left} = T_{\rm right} = T_w$). 

In order to simulate the Couette flow, we employ the LB models presented in 
Ref.~\cite{ambrus16jcp}. While these models were used in Ref.~\cite{ambrus16jcp} for the 
simulation of the $2D$ Couette flow in the Boltzmann-BGK model, their applicability 
to the present problem is immediate since the extra momentum space degree of freedom $p_z$, 
which is perpendicular to the flow direction and to the normal to the walls, can 
be eliminated by introducing the reduced distributions $\phi$ and $\chi$, as discussed below.
It was shown in Ref.~\cite{ambrus16jcp} that, for the simulation 
of flows at non-negligible values of ${\rm Kn}$, the half-range Gauss-Hermite quadrature should be 
used on the axis perpendicular to the walls (the $x$ axis), while a relatively low-order full-range 
Gauss-Hermite quadrature is adequate for the direction parallel to the walls.
The resulting models are denoted by ${\rm HHLB}(N_x; Q_x) \times {\rm HLB}
(N_y; Q_y)$,
where $N_x$ and $N_y$ are the orders of the polynomial 
expansions of the equilibrium distribution
along the $x$ and $y$ axes, respectively.
$Q_x$ and $Q_y$ denote the quadrature orders employed on the
$x$ semiaxes and on the full $y$ axis of the momentum space. 
The resulting velocity set comprises $2Q_x \times Q_y $ vectors.

In Ref.~\cite{ambrus16jcp} it was shown, for a two-dimensional mixed quadrature LB model,
that the value of $Q_x$ required to obtain accurate simulation
results depends on ${\rm Kn}$ and on the wall velocity $u_w$. 
Furthermore, it was also shown that, in the particular case of the 
BGK implementation of the collision term, the simulations employing 
$Q_y \ge 4$ at fixed $Q_x$ produced identical results. In this section, 
a similar inequality will be derived in the case of the Shakhov model.

Since the Couette flow is completely homogeneous along the directions which are parallel to the walls
(i.e.~the $y$ and $z$ directions), Eq.~\eqref{eq:boltz} reduces to:
\begin{equation}
 \partial_t f + \frac{p_x}{m} \partial_x f = -\frac{1}{\tau}\left[f - \feq(1+\mathbb{S})\right].
 \label{eq:boltz_x}
\end{equation}
The $p_z$ degree of freedom of the momentum space can be eliminated by integrating Eq.~\eqref{eq:boltz_x} with 
respect to $p_z$. Defining the reduced distributions 
$\phi\equiv\phi(x,p_x, p_y, t)$ and $\chi\equiv\chi(x,p_x, p_y, t)$ via:
\begin{equation}
 \phi = \int_{-\infty}^\infty dp_z\, f, \qquad 
 \chi = \int_{-\infty}^\infty dp_z\, f \, \frac{p_z^2}{m},\label{eq:reduced_def}
\end{equation}
the following two equations are obtained:
\begin{equation}
 \partial_t 
 \begin{pmatrix}
  \phi\\ \chi %\rule{0mm}{7mm}
 \end{pmatrix} + \frac{p_{x}}{m} \partial_x 
 \begin{pmatrix}
  \phi\\ \chi%\rule{0mm}{7mm}
 \end{pmatrix} = -\frac{1}{\tau} 
 \begin{pmatrix}
  \phi - \phi^{\rm (eq)} (1 + \mathbb{S}_\phi) \\ 
  \chi - \chi^{\rm (eq)} (1 + \mathbb{S}_\chi) %\rule{0mm}{7mm}
 \end{pmatrix},\label{eq:boltz_red}
\end{equation}
where $\chi^{\rm (eq)} = T \phi^{\rm (eq)}$ and $\phi^{\rm (eq)}$ is just the 2D Maxwell-Boltzmann distribution ($\alpha \in \{x,y\}$):
\begin{equation}
 \phi^{\rm (eq)} = n g_x g_y, \qquad g_\alpha = \frac{1}{\sqrt{2\pi m T}} 
 \exp\left[-\frac{(p_\alpha -mu_\alpha)^2}{2m T}\right].
 \label{eq:phieq}
\end{equation}
The terms $\mathbb{S}_\phi$ and $\mathbb{S}_\chi$ are given by:
\begin{align}
 \mathbb{S}_\phi =& \int_{-\infty}^\infty \frac{dp_z}{\sqrt{2\pi m T}} e^{-p_z^2 / 2mT} \, \mathbb{S} 
 \nonumber\\
 =& \frac{1-{\rm Pr}}{nT^2}
\left(\frac{\xi_x^{2} + \xi_y^2}{5mT}-\frac{4}{5}\right) (\xi_x q_x + \xi_y q_y),\nonumber\\
 \mathbb{S}_\chi =& \int_{-\infty}^\infty \frac{dp_z}{\sqrt{2\pi m T}} e^{-p_z^2 / 2mT} 
 \frac{p_z^2}{m T}\, \mathbb{S} \nonumber\\
 =& \frac{1-{\rm Pr}}{nT^2}
 \left(\frac{\xi_x^{2} + \xi_y^2}{5mT}-\frac{2}{5}\right) (\xi_x q_x + \xi_y q_y).
 \label{eq:sred_def}
\end{align}

Let us now expand $\phi$ and $\phi^{\rm (eq)}(1+\mathbb{S}_\phi)$ with respect to the full-range 
Hermite polynomials on the $y$ axis, as follows \cite{hildebrand}:
\begin{align}
 \phi =& \frac{\omega(\wp_y)}{p_{0,y}} \sum_{\ell = 0}^\infty \frac{1}{\ell!} \mathcal{F}_{\ell}(x, p_x, t) 
 H_\ell(\wp_y), \nonumber\\
 \phi^{\rm (eq)} (1+\mathbb{S}_\phi) =& \frac{\omega(\wp_y)}{p_{0,y}}
 \sum_{\ell = 0}^\infty \frac{1}{\ell!} \mathcal{F}^{\mathbb{S}_\phi}_{\ell}(x, p_x, t) H_\ell(\wp_y),
 \label{eq:fyz}
\end{align}
where $\wp_y \equiv p_y / p_{0,y}$ represents the particle momentum 
along the $y$ axis in units of an arbitrary momentum scale $p_{0,y}$ 
(we only consider $p_{0,y} = 1$ in this paper), while
the weight function $\omega(\wp_y)$ for the 
full-range Hermite polynomials is
\begin{equation}
 \omega(\wp_y) = \frac{1}{\sqrt{2\pi}} e^{-\wp_y^2/2}.\label{eq:H_omega}
\end{equation}
Furthermore, the coefficients $\mathcal{F}_{\ell}$ and $\mathcal{F}^{\mathbb{S}_\phi}_{\ell}$ can be computed as:
\begin{align}
 \mathcal{F}_{\ell} =& \int_{-\infty}^\infty dp_y \, \phi\, H_\ell(\wp_y), \nonumber\\
 \mathcal{F}^{\mathbb{S}_\phi}_{\ell} =& \int_{-\infty}^\infty dp_y \, 
 \phi^{\rm (eq)} (1+\mathbb{S}_\phi)\, H_\ell(\wp_y).
 \label{eq:Fmn_def}
\end{align}
The compatibility between Eqs.~\eqref{eq:fyz} and \eqref{eq:Fmn_def} is ensured by the orthogonality relation obeyed by 
the Hermite polynomials \cite{nist}:
\begin{equation}
 \int_{-\infty}^\infty d\wp_y \, \omega(\wp_y) H_\ell(\wp_y) H_{\ell'}(\wp_y) = \ell! \,\delta_{\ell\ell'}.
\label{eq:H_ortho}
\end{equation}

Substituting \eqref{eq:fyz} into the Boltzmann equation \eqref{eq:boltz_x} and projecting on 
the space of Hermite polynomials gives:
\begin{equation}
 \partial_t \mathcal{F}_{\ell} + \frac{p_x}{m} \partial_x \mathcal{F}_{\ell} = 
 -\frac{1}{\tau}\left(\mathcal{F}_{\ell} - \mathcal{F}^{\mathbb{S}_\phi}_{\ell}\right),
\end{equation}
where, as before, $\mathcal{F}_{\ell}$ and $\mathcal{F}^{\mathbb{S}_\phi}_{\ell}$ depend on $x$, $p_x$ and $t$.
Since $\phi^{\rm (eq)}$ depends on
$n$, $\vu$ and $T$, while $\mathbb{S}_\phi$ depends on $\vq$, the coefficients 
$\mathcal{F}^{\mathbb{S}_\phi}_{\ell}$ always involve the coefficients $\mathcal{F}_{\ell'}$ of orders 
$0 \le \ell' \le 3$. 
In this paper, only the moments of $f$ up to $\bm{q}$ are tracked.
These moments can be expressed 
with respect to the coefficients $\mathcal{F}_{\ell}$ with $0 \le \ell \le 3$.
In our implementation, 
we consider the expansion \eqref{eq:phieq} of $\phi^{\rm (eq)}$ through a direct 
product procedure, by separately expanding the factors $g_x$ and $g_y$. 
More precisely, $g_x$ is expanded with respect to the half-range Hermite polynomials and
$g_y$ is expanded with respect to the full-range Hermite polynomials up 
to order $N_y = Q_y - 1$ \cite{ambrus16jcp,ambrus16jocs}. Since 
$\mathbb{S}_\phi$ is a polynomial of the third order in $\vp$, 
the evolution of $\vq$ requires the recovery 
of the moments of $\phi^{\rm (eq)}$ of order $3 + 3 = 6$ on each axis. 
Such moments can be exactly recovered when $N_y \ge 6$.

The above analysis shows that, when the Shakhov collision term
is employed, the numerical results for the 
$3D$ Couette flow considered in this paper when $Q_y > 7$ must coincide with those obtained with $Q_y = 7$.
Furthermore, as discussed in Ref.~\cite{ambrus16jcp}, the accurate simulation of flows at large values of 
${\rm Kn}$ requires the increase of the quadrature $Q_x$. 
It is worth mentioning that if ${\rm Pr} = 1$, the Shakhov term 
$\mathbb{S}$ in Eq.~\eqref{eq:sdef} vanishes and the LB models 
with $Q_y \ge 4$ give identical 
results, as discussed in Ref.~\cite{ambrus16jcp}.

\section{Low Mach number validation}\label{sec:lowmach}

\begin{figure*}
\begin{center}
\begin{tabular}{cc}
 \includegraphics[width=0.475\linewidth]{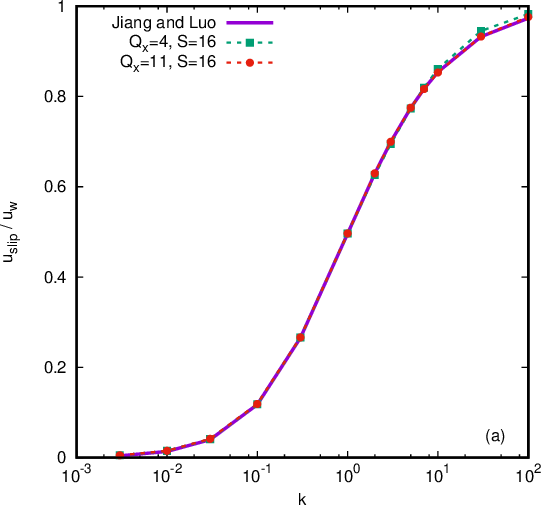} &
 \includegraphics[width=0.475\linewidth]{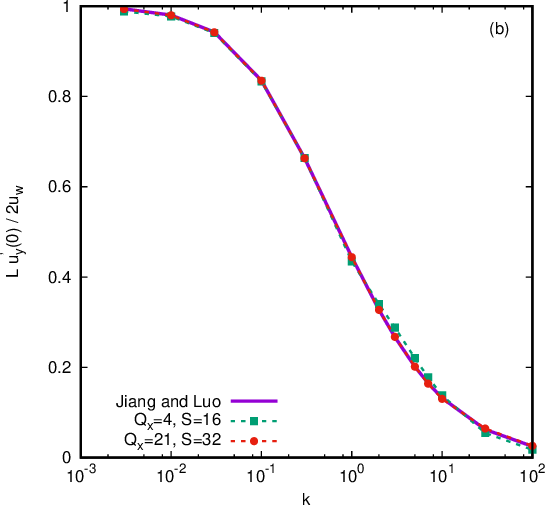} \\
 \includegraphics[width=0.475\linewidth]{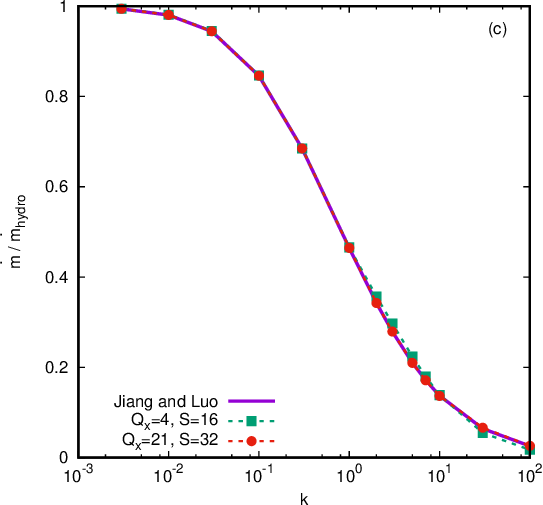} &
 \includegraphics[width=0.475\linewidth]{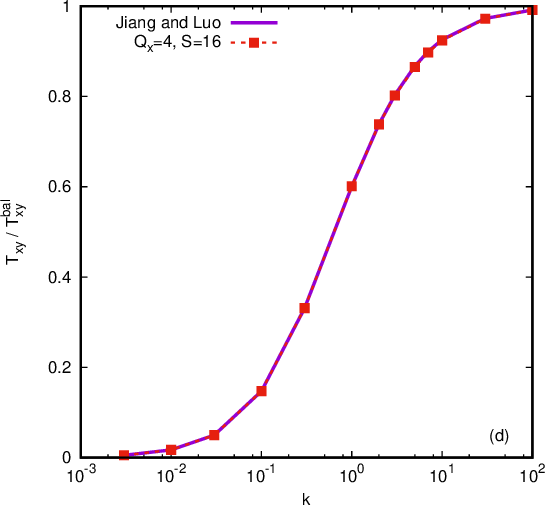} 
\end{tabular}
\caption{Comparison between the results obtained 
using our models for various quadrature orders $Q_x$ and 
number of grid points $S$ (dashed lines and points) and the 
benchmark data on the solution 
of the linearized Boltzmann equation reported by Jiang and Luo 
in Ref.~\cite{jiang16} (continuous lines), shown with respect 
to the Knudsen number $k$ defined in Eq.~(\ref{eq:knLuo}), for
(a) $u_{\rm slip} / u_w$; 
(b) $L u'_y(0) / 2 u_w$; 
(c) $\dot{m} / \dot{m}_{\rm hydro}$; and 
(d) $T_{xy} / T_{xy}^{\rm bal}$.
}
\label{fig:slip_res}
\end{center}
\end{figure*}

\begin{figure*}
\begin{center}
\begin{tabular}{cc}
 \includegraphics[width=0.475\linewidth]{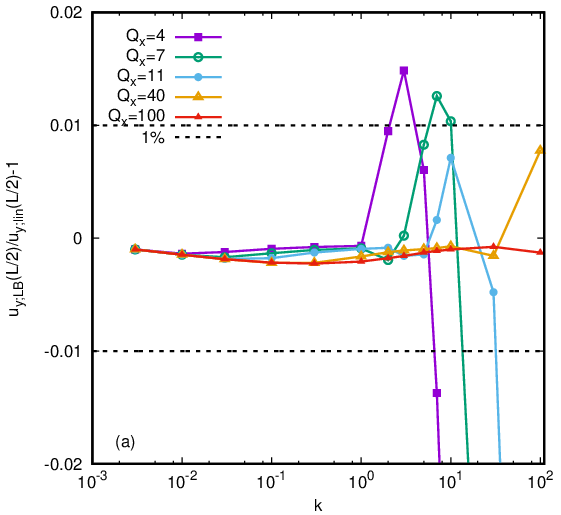} &
 \includegraphics[width=0.475\linewidth]{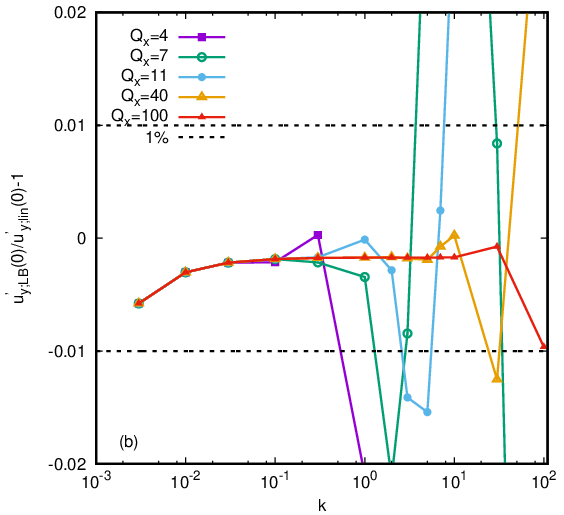} \\
 \includegraphics[width=0.475\linewidth]{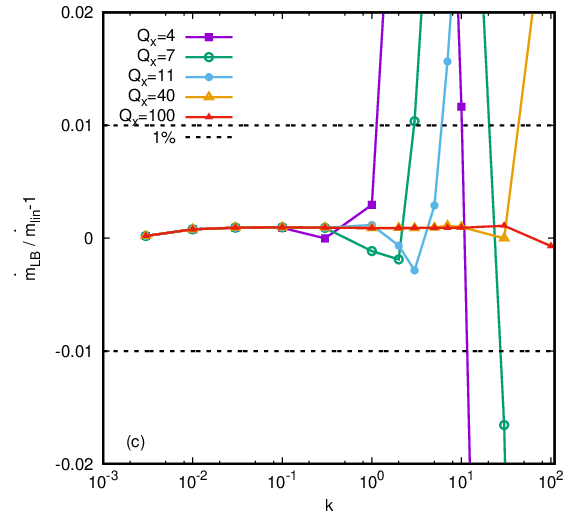} &
 \includegraphics[width=0.475\linewidth]{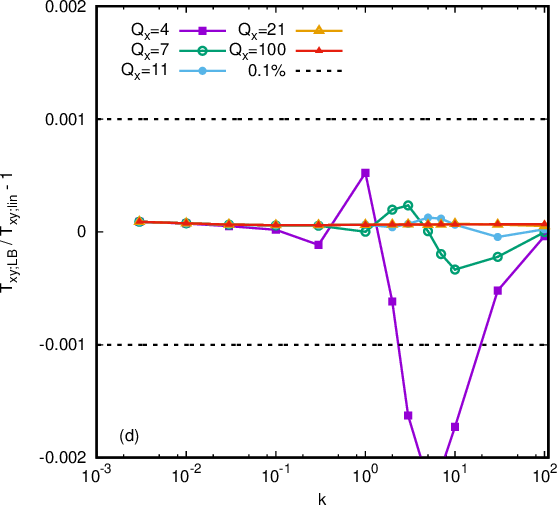} 
\end{tabular}
\caption{Effect of the quadrature order on the 
relative error $A_{\rm LB} / A_{\rm lin} - 1$
between our numerical results and the 
benchmark data reported by Jiang and Luo in Ref.~\cite{jiang16},
shown with respect to the Knudsen number $k$
defined in Eq.~(\ref{eq:knLuo}), where the quantity $A$ stands for 
(a) $u_{y}(L/2)$; (b) $u'_y(0)$; (c) $\dot{m}$; and 
(d) $T_{xy}$.
}
\label{fig:slip_err}
\end{center}
\end{figure*}

\begin{table}
\begin{tabular}{|l|r|r|r|}
\hline
 $k$ & $N_x$ & $Q_x$ & $S$\\\hline
 $0.03$ & $3$ & $4$ & $64$ \\
 $0.1$ & $6$ & $7$ & $32$ \\
 $1$ & $6$ & $21$ & $32$ \\
 $10$ & $6$ & $80$ & $32$\\\hline
\end{tabular}
\caption{Simulation parameters for the results shown 
in Fig.~\ref{fig:luo}(a).}
\label{tab:luo}
\end{table}

\begin{figure}
\begin{center}
\begin{tabular}{c}
 \includegraphics[width=0.95\linewidth]{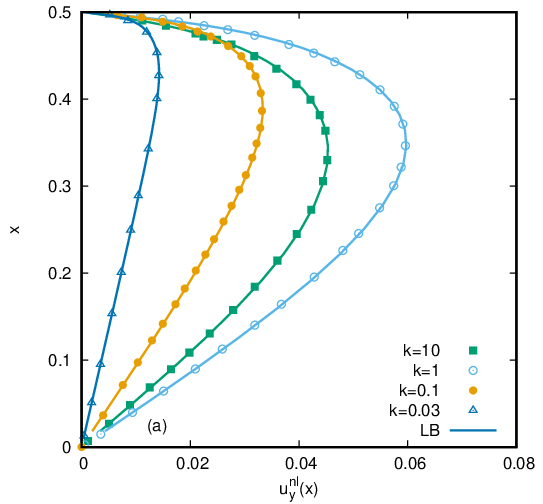}\\
 \includegraphics[width=0.95\linewidth]{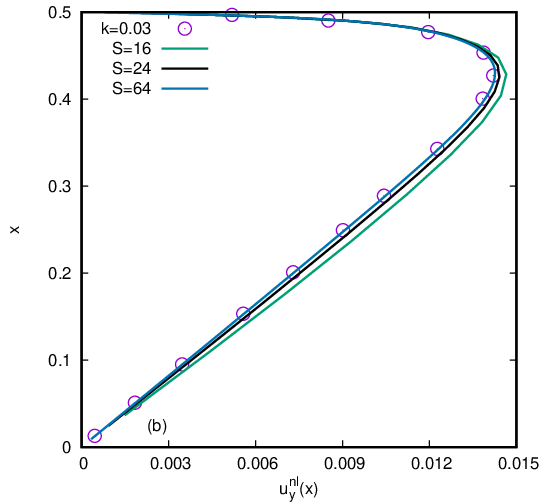} 
\end{tabular}
\caption{Comparison between our numerical results and those 
reported by Li et al in Ref.~\cite{li15} for the nonlinear part 
of the velocity profile \eqref{eq:uy_nl}:
(a) for various values of the Knudsen number
$k \in \{0.03,\, 0.1\,, 1,\, 10\}$ (the simulation parameters 
are summarized in Table~\ref{tab:luo});
(b) for $k = 0.03$, $Q_x = 4$, and the number of grid points 
$S \in \{16,\, 24,\, 64\}$.
The Knudsen number $k$ is defined in Eq.~(\ref{eq:knLuo}).
}
\label{fig:luo}
\end{center}
\end{figure}

For low Mach number flows, the Boltzmann equation can be approximated via 
its linearized form, 
which was tackled by many authors using numerical simulations 
\cite{cercignani69,kim08,sharipov09,sharipov16,wu2017} or semi-analytic 
methods \cite{sone07,yap12,li15,jiang16}. In order to approach the 
assumptions of the 
linearized regime, the simulations presented in this section are performed with 
$u_w = 10^{-5}$,
such that the temperature and density profiles remain nearly constant 
throughout the channel. At higher values 
of $u_w$, deviations from the benchmark results can be expected, 
since the profile of $u_y /u_w$ does not scale perfectly with $u_w$, as also 
demonstrated 
in Ref.~\cite{rogozin16}. At $u_w = 10^{-5}$, a quadrature order $Q_y = 2$ is 
sufficient  for the direction pointing along the flow.

Since the value of ${\rm Pr}$ does not influence the results in this regime, 
the simulations are performed with ${\rm Pr} = 1$ (the BGK approximation).
Furthermore, since the linearized regime analysis is oblivious of the 
number of degrees of freedom of the particle constituents, our simulations 
are performed only at the level of the $\phi$ function, as follows:
\begin{equation}
 \partial_t \phi + \frac{p_x}{m} \partial_x \phi = -\frac{1}{\tau}[\phi - \phi^{\rm (eq)}],
\end{equation}
where $\phi^{\rm (eq)}$ is given in Eq.~\eqref{eq:phieq} with the 
temperature obtained as $T = \frac{1}{2n}(T_{xx} + T_{yy})$.
The relaxation time is implemented through:
\begin{equation}
 \tau = \frac{k}{\sqrt{2}},
 \label{eq:knLuo}
\end{equation}
where $k$ is interpreted as the Knudsen number \cite{sone07,li15,jiang16}.

The numerical results presented in this section were 
obtained with the ${\rm HHLB}(N_x; Q_x) \times {\rm HLB}(1;2)$ 
models, for various values of $Q_x$. For consistency with the 
subsequent numerical results sections, the expansion order $N_x$
was set to $N_x = {\rm min}(6, Q_x - 1)$. The spatial domain 
was discretized using $S$ nodes, stretched according to 
Eq.~\eqref{eq:etas} with $A = 0.98$. More details on the 
numerical scheme are provided in the Appendix.

In the first part, we consider a comparison with the benchmark data 
obtained by Jiang and Luo in Ref.~\cite{jiang16} for the following 
four quantities: the slip velocity $u_{\rm slip}$, the velocity 
derivative at the channel center $u'(0)$, the mass flow rate 
$\dot{m}$, and the non-diagonal stress $T_{xy}$, which are 
introduced below. 

The slip velocity is obtained by subtracting the fluid velocity 
at the wall $u_y(L/2)$
from the wall velocity:
\begin{equation}
 u_{\rm slip} = u_w - u_y(L/2),\label{eq:uslip}
\end{equation}
where $u_y(L/2)$ is obtained by quadratic extrapolation from the 
fluid nodes with indices $S$, $S-1$ and $S-2$
($S$ is the index of the last node inside the fluid domain
and $u_{y;i}$ is the $y$ component of the fluid velocity
in node $i,\,1\leq i \leq S$):
\begin{multline}
 u_y(L/2) = \frac{(L/2-x_{S-1}) (L/2-x_{S-2}) u_{y;S}}
 {(x_S - x_{S-1})(x_S - x_{S-2})} \\+ 
 \frac{(L/2-x_{S}) (L/2-x_{S-2}) u_{y;S-1}}
 {(x_{S-1} - x_{S})(x_{S-1} - x_{S-2})} \\+ 
 \frac{(L/2-x_{S}) (L/2-x_{S-1}) u_{y;S-2}}
 {(x_{S-2} - x_{S})(x_{S-2} - x_{S-1})}.\label{eq:uslip_extra}
\end{multline}
This expression of $ u_y(L/2)$ is third order accurate with respect 
to the spacing $\delta \eta$ of the parameter of the stretched grid,
which is defined in the Appendix.

The velocity derivative at the channel center is obtained using the following 
three-point formula:
\begin{multline}
 u_y'(0) = \frac{x_2^2 x_3^2 u_{y;1}}{x_1 (x_1^2 - x_2^2)(x_1^2 - x_3^2)} \\+
 \frac{x_1^2 x_3^2 u_{y;2}}{x_2 (x_2^2 - x_1^2)(x_2^2 - x_3^2)} +
 \frac{x_1^2 x_2^2 u_{y;3}}{x_3 (x_3^2 - x_1^2)(x_3^2 - x_2^2)},
 \label{eq:ud0_extra}
\end{multline}
which is sixth order accurate if we take into account the antisymmetry of the 
velocity profile with respect to the channel center. 

The half-channel mass flow rate is obtained using the rectangle integration method:
\begin{align}
 \dot{m} = \int_0^{L/2} dx\, \rho u_y =& 
 \frac{L}{2A} \int_0^{{\rm arctanh}\, A} \frac{d\eta}{\cosh^2\eta}\rho u_y \nonumber\\
 \simeq& \frac{L\, {\rm arctanh}\, A}{2AS} \sum_{s= 1}^{S}
 \frac{\rho_s u_{y,s}}{\cosh^2 \eta_s},\label{eq:mflow}
\end{align}
which is second order accurate with respect to $\delta \eta$. 
In the hydrodynamic regime, when $u_y^{\rm hydro} = 2 u_w x / L$ 
\cite{kundu15,rieutord15}, the half-channel mass flow rate 
is given by:
\begin{equation}
 \dot{m}_{\rm hydro} = \int_0^{L/2} dx\, \rho u_y^{\rm hydro} 
 = \frac{1}{4} \rho u_w L.\label{eq:mflow_hydro}
\end{equation}

Finally, the non-diagonal stress $T_{xy}$ is obtained by averaging 
$T_{xy}$ over the half-channel, using the equivalent of Eq.~\eqref{eq:mflow}:
\begin{equation}
 T_{xy} = \frac{{\rm arctanh}\, A}{AS} \sum_{s= 1}^{S}
 \frac{T_{xy;s}}{\cosh^2 \eta_s}.\label{eq:Txy}
\end{equation}
In the ballistic regime, $T_{xy}$ is given by \cite{sone07,ambrus16jcp}:
\begin{equation}
 T_{xy}^{\rm bal} = -\rho u_w \sqrt{\frac{2 T_w}{m \pi}}.
 \label{eq:Txy_bal}
\end{equation}

In Fig.~\ref{fig:slip_res}, the results obtained with our mixed quadrature 
LB models are compared with those reported by Jiang and Luo 
in Table~2  of Ref.~\cite{jiang16}, for
(a) the relative slip velocity $u_{\rm slip} / u_w$,
(b) the normalized velocity derivative at the channel center 
$L u'_y(0) / 2u_w$, (c) the normalized mass flow rate 
$\dot{m} / \dot{m}_{\rm hydro}$ and 
(d) the normalized non-diagonal stress $T_{xy} / T_{xy}^{\rm bal}$.
It can be seen that very reasonable agreement is obtained with 
the ${\rm HHLB}(3;4) \times {\rm HLB}(1;2)$ model on a grid 
with $S = 16$ nodes. The small discrepancies seen in
Fig.~\ref{fig:slip_res}(b)
at small values of $k$ 
are removed by doubling the grid points, 
while the discrepancies observed at large values of $k$ in
Figs.~\ref{fig:slip_res}(a-c)  can 
be removed by increasing the quadrature order $Q_x$ of the 
half-range Gauss-Hermite quadrature. 

The comparison shown in Fig.~\ref{fig:slip_res} allows a qualitative 
assessment to be made at the level of absolute differences 
between our results and the benchmark results. We now discuss the 
relative error $\varepsilon(A)$, which is defined for a quantity $A$ as 
\begin{equation}
 \varepsilon(A) = \frac{A_{\rm LB}}{A_{\rm lin}} - 1,
\end{equation}
where $A_{\rm LB}$ and $A_{\rm lin}$ are the values of $A$ obtained 
using our models and those reported in Ref.~\cite{jiang16}, respectively.
The use of $\varepsilon(A)$ augments the differences between 
our LB results and the benchmark data in the regions where
$A$ is small.
In Fig.~\ref{fig:slip_err}(a-d),
 the relative errors $\varepsilon(A)$, computed for 
$A\in \{u_y(L/2), u'_y(0), \dot{m}, T_{xy}\}$,  are represented 
with respect to $k$ \eqref{eq:knLuo} for various quadrature orders. 
The dashed lines in panels (a--c) and (d) 
indicate the $1\%$ and $0.1\%$ relative error thresholds, respectively. 
It can be seen that 
for the quantities (a) $u_y(L/2)$, (b) $u'_y(0)$, and
(c) $\dot{m}$, the relative error at high values of $k$ can be 
decreased below $1\%$ only when
high quadrature orders $Q_x$ are employed.
This is because these three quantities go to $0$ as $k \rightarrow \infty$ 
and hence the absolute error must decrease significantly in order to 
achieve the $1\%$ threshold for the relative error. By contrast, the relative error 
in $T_{xy}$, shown in Fig.~\ref{fig:slip_err}(d),
 is well below $1\%$ even when $Q_x = 4$. For this
 quantity, the relative error becomes less that $0.1\%$ when $Q_x \ge 7$.
All results presented in Fig.~\ref{fig:slip_err} were 
obtained on a $1D$ grid with $S = 16$ points,
stretched according to Eq.~\eqref{eq:stretch} with $A = 0.98$,
 using the models ${\rm HHLB}(N_x; Q_x) \times {\rm HLB}(1; 2)$ 
of various quadrature orders $Q_x$ and $N_x = {\rm min}(6, Q_x - 1)$.

In the second and final part, a comparison between our results and those reported in 
Ref.~\cite{li15} for the nonlinear part of the velocity profile 
$u_y^{\rm nl}$ is considered. This nonlinear part refers to the 
departure of the solution of the kinetic equation from the straight
line profile predicted via the Navier-Stokes equations. 
The construction of $u_y^{\rm nl}$ is made by first obtaining a 
linear velocity profile which vanishes at the channel center and 
which attains the value predicted by the kinetic equation on the channel 
wall. This linear
profile can be regarded as the solution of the Navier-Stokes 
equations with the correct velocity slip taken into account.
Subtracting the velocity profile obtained by solving the kinetic equation 
gives a profile which vanishes, by construction, at the channel center and 
at the wall. This nonlinear velocity profile $u_y^{\rm nl}$
is defined as:
\begin{equation}
 u_y^{\rm nl} = \frac{2x}{L} - \frac{u_y(x)}{u_y(L/2)}.
 \label{eq:uy_nl}
\end{equation}
It can be seen in Fig.~\ref{fig:luo}(a) that 
the resulting profiles cancel at $x = 0$ and $x = L/2$, while 
reaching a maximum value inside the channel. The value of this 
maximum increases with $k$ 
up to a maximum value, 
and decreases afterwards as $k \rightarrow \infty$, when the velocity 
profile is trivially a straight line: 
$u_y^{\rm bal}(x) = 0$. Our results are in excellent agreement 
with those reported in Ref.~\cite{li15} and were obtained using
the models ${\rm HHLB}(N_x; Q_x) \times {\rm HLB}(1; 2)$, with 
$N_x = {\rm min}(Q_x - 1, 6)$, while 
$Q_x = 4, 7, 21$ and $80$ for $k = 0.03$, $0.1$, $1$ and $10$, respectively. 
In order to maintain a good accuracy,
the number of grid points had to be increased  to $S = 64$ for $k = 0.03$ and $S = 32$ for 
$k \in \{0.1,\,1,\,10\}$. These simulation parameters are summarized 
in Table~\ref{tab:luo}. For $k = 0.03$, the approach of our LB results 
towards the benchmark solution as the grid is successively 
refined can be seen in Fig.~\ref{fig:luo}(b).

\section{Comparison with DSMC: Maxwell molecules} \label{sec:mw}

\begin{table}
\begin{center}
\begin{tabular}{|c||c|c|c|}
 \hline ${\rm Kn}$ & $\widetilde{l}_{\rm ref}$ & $\widetilde{t}_{\rm ref}$ \\\hline\hline 
 $0.01$ & $0.8833 \,{\rm m}$ & $3.71\,{\rm ms}$ \\\hline
 $0.05$ & $0.17666 \,{\rm m}$ & $0.741 \,{\rm ms}$ \\\hline 
 $0.1$ & $0.08833 \,{\rm m}$ & $0.371\,{\rm ms}$ \\\hline
 $0.25$ & $0.035332 \,{\rm m}$ & $0.148 \,{\rm ms}$ \\\hline 
 $0.5$ & $0.017666 \,{\rm m}$ & $74.1 \, \mu{\rm s}$ \\\hline 
 $1.0$ & $0.008833 \,{\rm m}$ & $37.1 \, \mu{\rm s}$ \\\hline
\end{tabular}
\end{center}
\caption{Reference values for the length and time for various values of ${\rm Kn}$,
in the context of the simulations discussed in Sec.~\ref{sec:mw}.
\label{tab:ref}
}
\end{table}

\begin{figure*}
\begin{center}
\begin{tabular}{ccc}
 \includegraphics[width=0.32\linewidth]{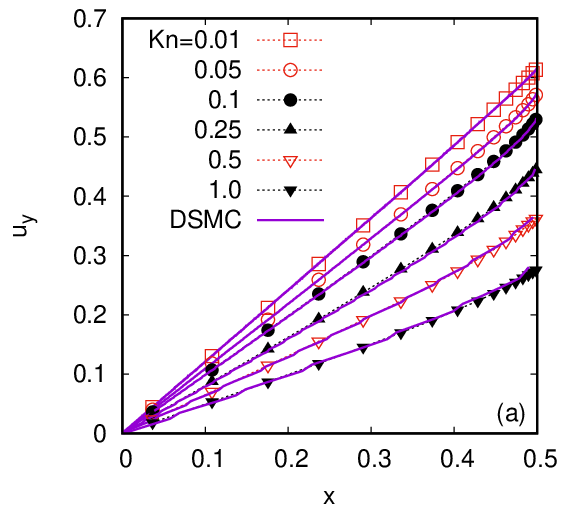} &
 \includegraphics[width=0.32\linewidth]{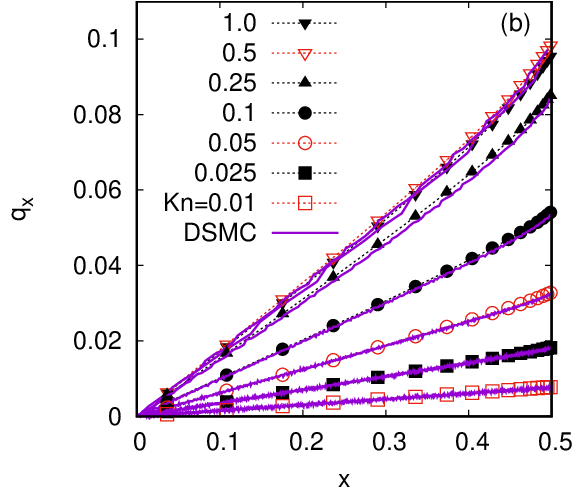} &
 \includegraphics[width=0.32\linewidth]{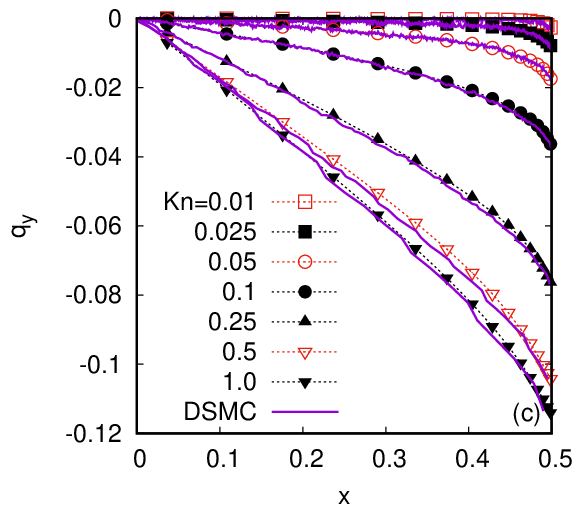} \\
 \includegraphics[width=0.32\linewidth]{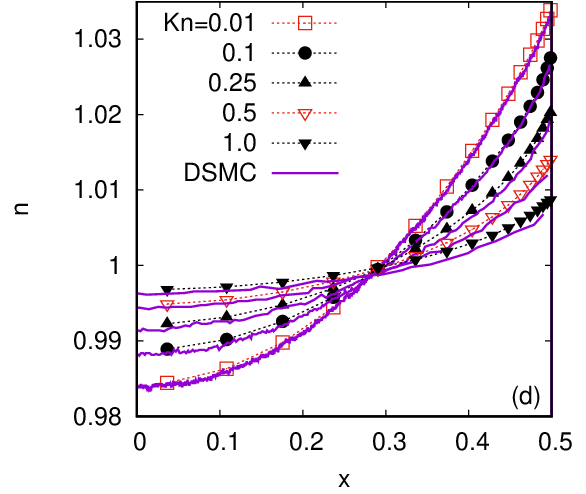} &
 \includegraphics[width=0.32\linewidth]{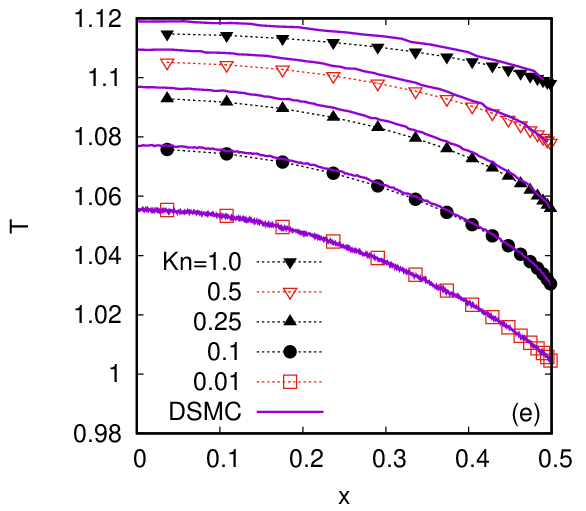} &
 \includegraphics[width=0.32\linewidth]{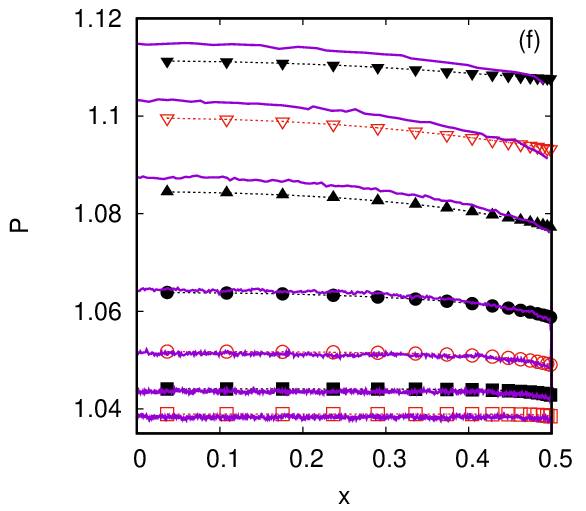} \\
 \includegraphics[width=0.32\linewidth]{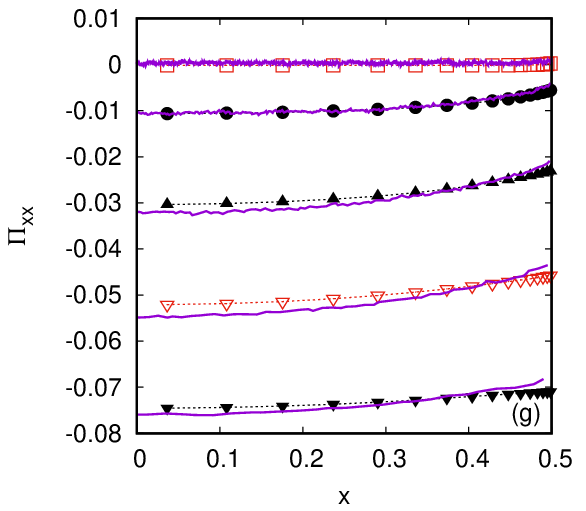} &
 \includegraphics[width=0.32\linewidth]{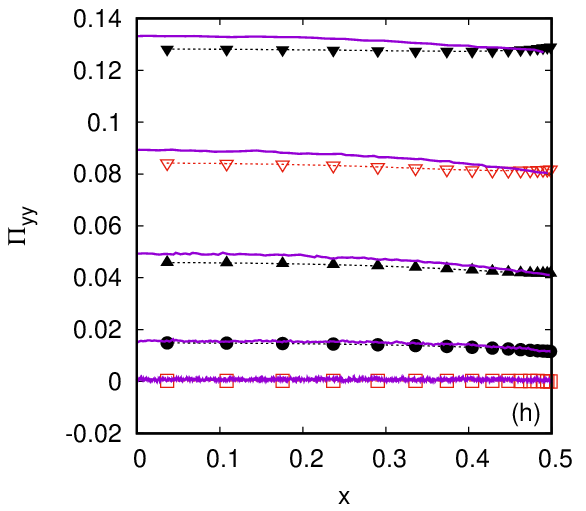} &
 \includegraphics[width=0.32\linewidth]{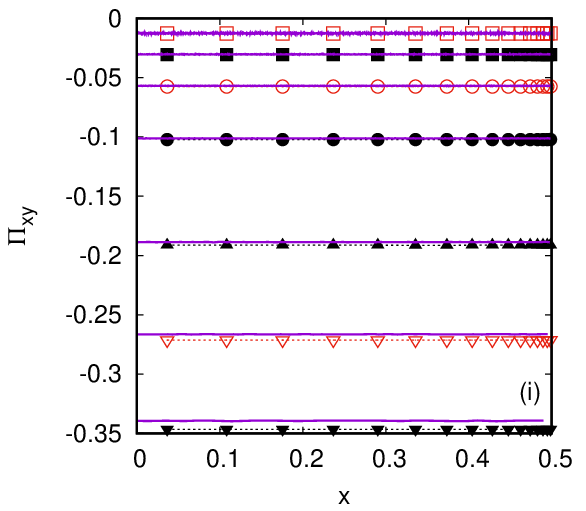}
\end{tabular}
\caption{Comparison of the LB results (dotted lines and points) and DSMC results 
from Refs.~\cite{struchtrup07,struchtrup08,schuetze03} (lines) for Maxwell molecules,
with $u_w \simeq 0.63$ and various values of ${\rm Kn}$. 
(a) velocity $u_y$; (b--c) components of the heat flux $q_i$; 
(d) density $n$; (e) temperature $T$; (f) pressure $P$; and
(g--i) components of the shear stress $\Pi_{ij} = T_{ij} - n T \delta_{ij}$.
}
\label{fig:mw-u063}
\end{center}
\end{figure*}

\begin{figure*}
\begin{center}
\begin{tabular}{ccc}
 \includegraphics[width=0.32\linewidth]{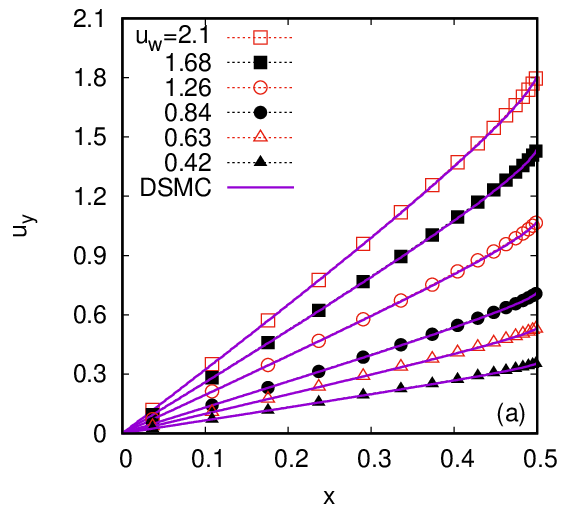} &
 \includegraphics[width=0.32\linewidth]{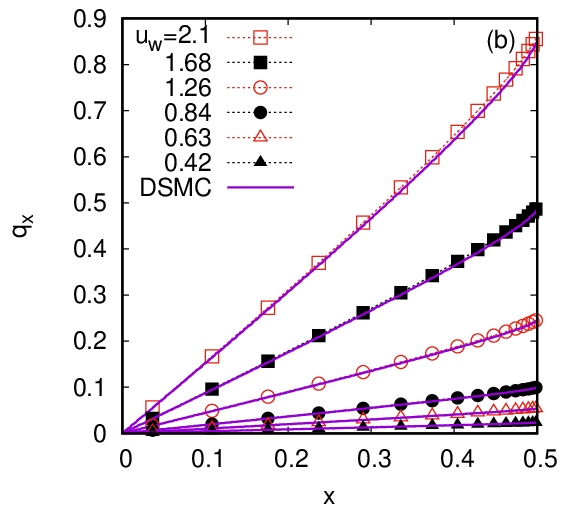} &
 \includegraphics[width=0.32\linewidth]{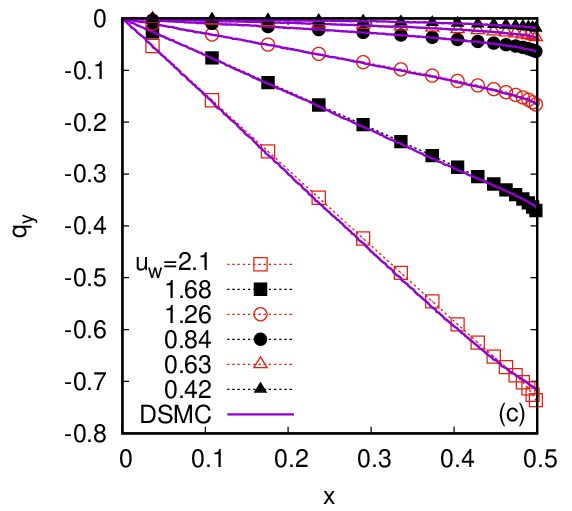} \\
 \includegraphics[width=0.32\linewidth]{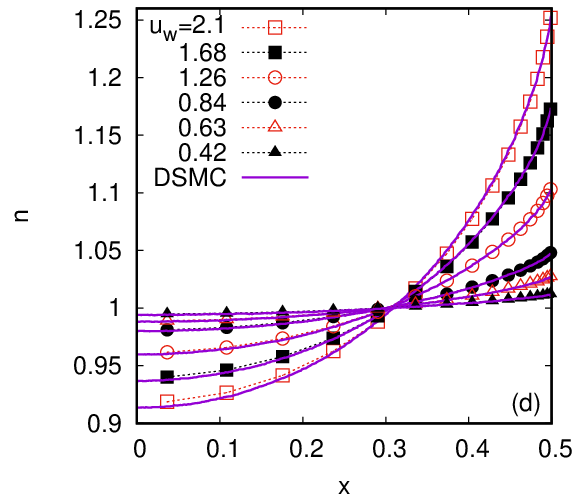} &
 \includegraphics[width=0.32\linewidth]{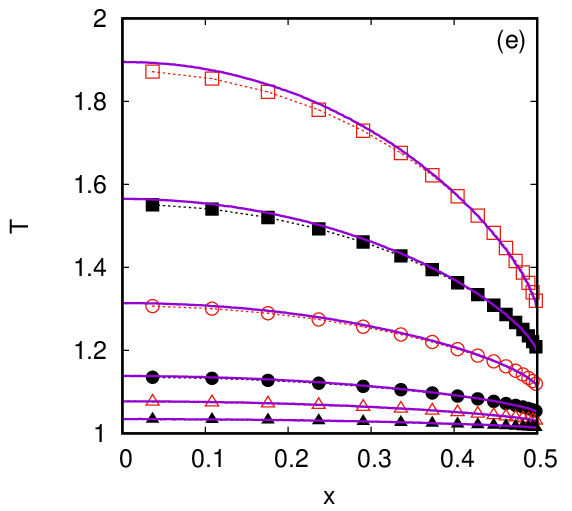} &
 \includegraphics[width=0.32\linewidth]{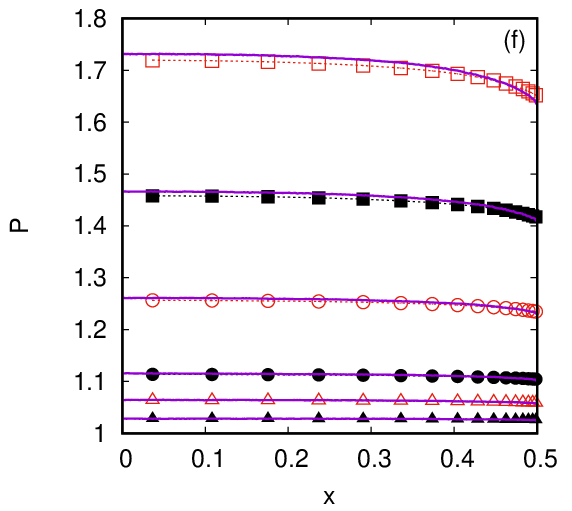} \\
 \includegraphics[width=0.32\linewidth]{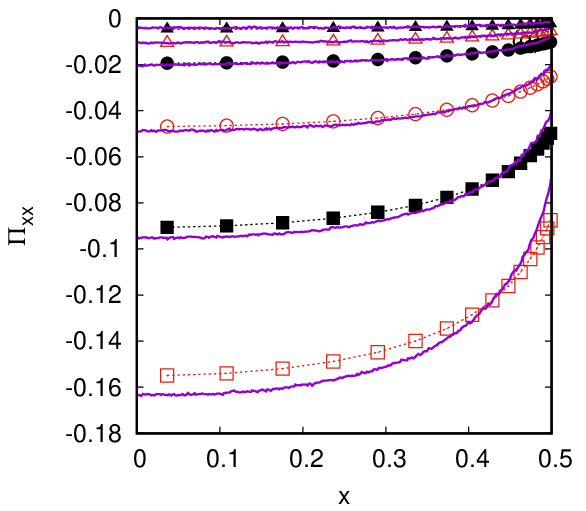} &
 \includegraphics[width=0.32\linewidth]{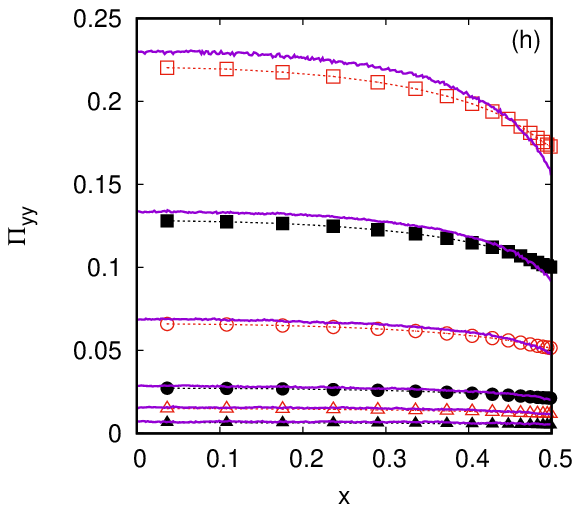} &
 \includegraphics[width=0.32\linewidth]{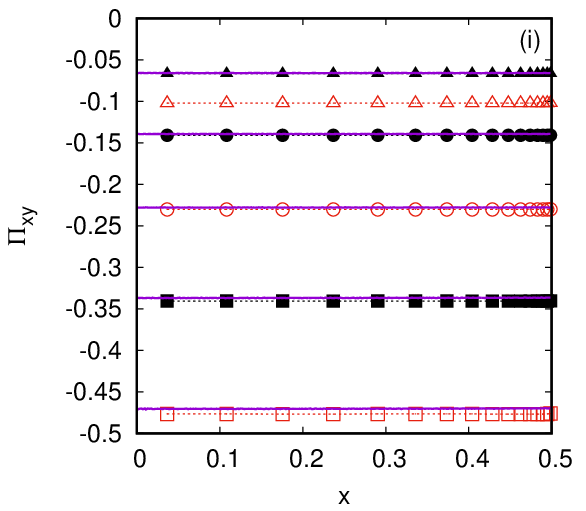}
\end{tabular}
\caption{Comparison of the LB results (dotted lines and points) and DSMC results 
from Refs.~\cite{struchtrup07,struchtrup08,schuetze03} (lines) for Maxwell molecules,
at ${\rm Kn} = 0.1$, for various values of the wall velocity $u_w$. 
(a) Velocity $u_y$; (b) transverse heat flux $q_x$; 
(c) longitudinal heat flux $q_y$; (d) density $n$; (e) temperature $T$; 
(f) pressure $P$; and (g--i) components of the viscous stress $\Pi_{ij} = T_{ij} - n T \delta_{ij}$.
}
\label{fig:mw-kn01}
\end{center}
\end{figure*}

\begin{figure*}
\begin{center}
\begin{tabular}{ccc}
 \includegraphics[width=0.32\linewidth]{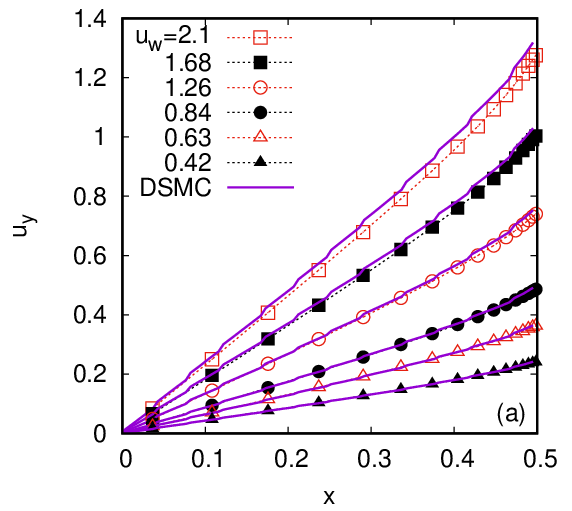} &
 \includegraphics[width=0.32\linewidth]{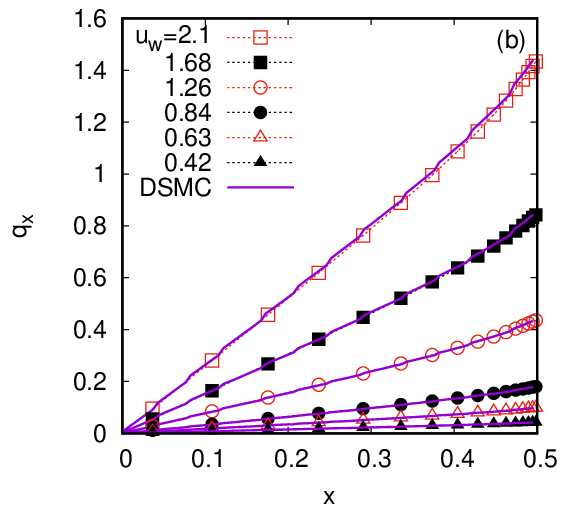} &
 \includegraphics[width=0.32\linewidth]{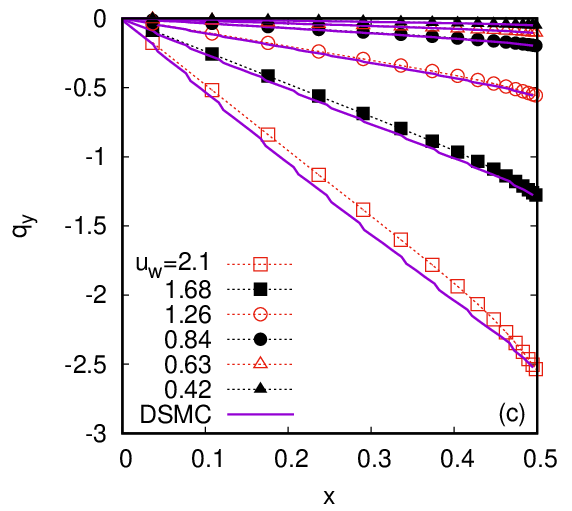} \\
 \includegraphics[width=0.32\linewidth]{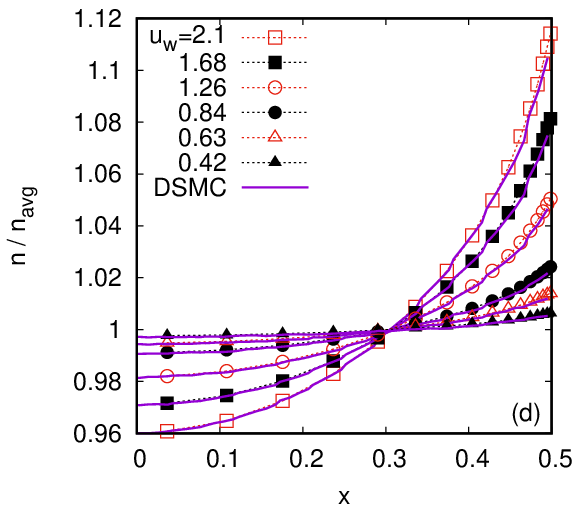} &
 \includegraphics[width=0.32\linewidth]{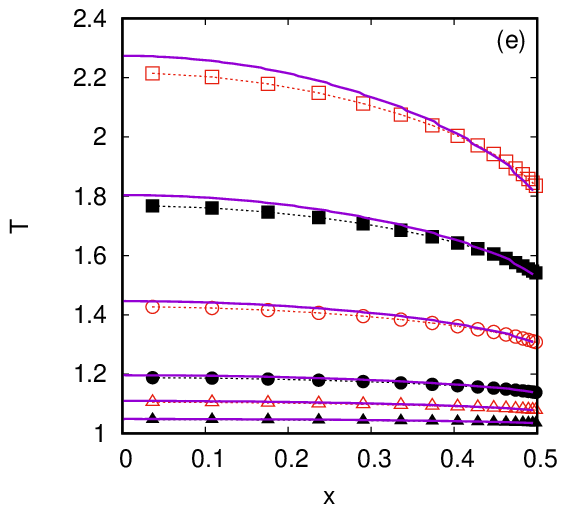} &
 \includegraphics[width=0.32\linewidth]{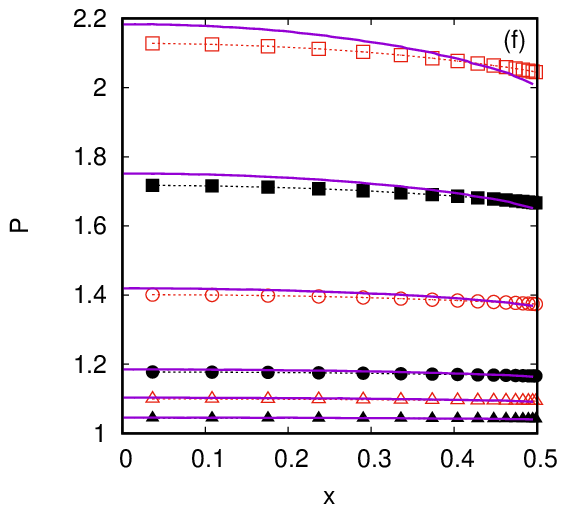} \\
 \includegraphics[width=0.32\linewidth]{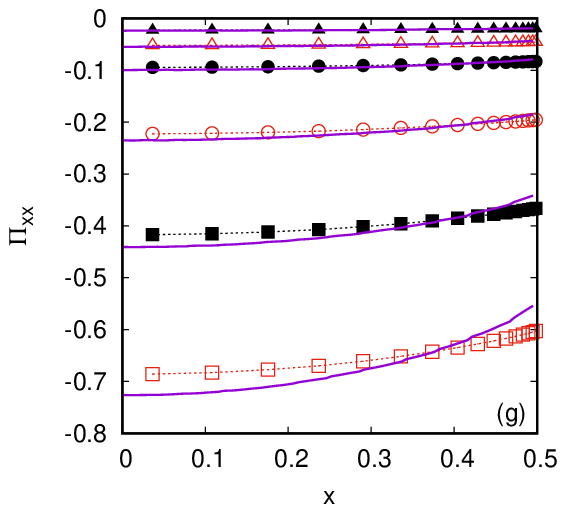} &
 \includegraphics[width=0.32\linewidth]{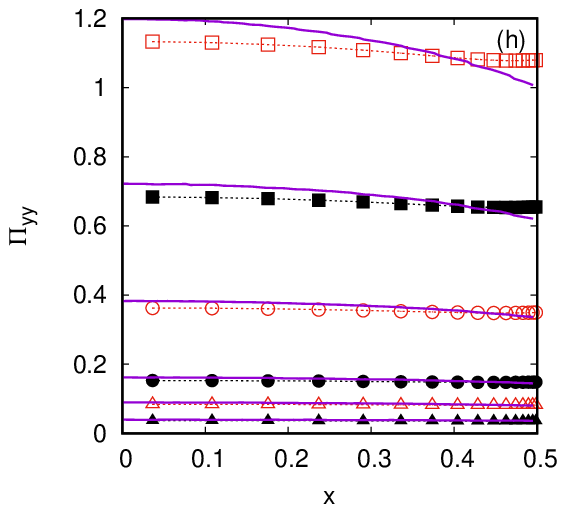} &
 \includegraphics[width=0.32\linewidth]{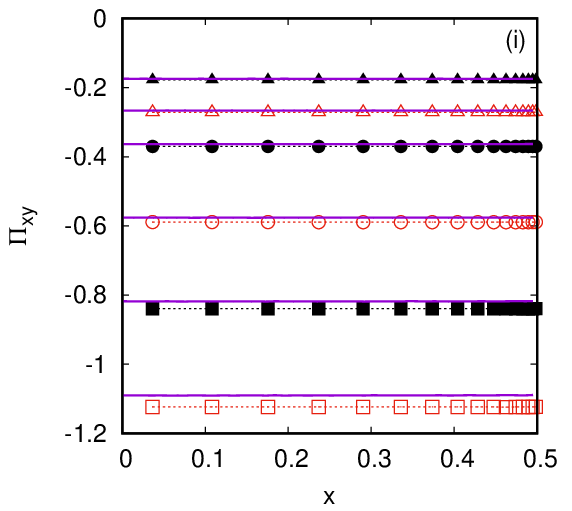}
\end{tabular}
\caption{Same as Fig.~\ref{fig:mw-kn01} for ${\rm Kn} = 0.5$.
}
\label{fig:mw-kn05}
\end{center}
\end{figure*}

We now benchmark our LB results against the direct simulation Monte Carlo 
(DSMC) results for Maxwell molecules reported by Struchtrup et al in 
Refs.~\cite{struchtrup07,struchtrup08,schuetze03}.
The working gas in these simulations is Argon, such that the reference 
mass is taken to be $\widetilde{m}_{\rm ref} = 
\widetilde{m}_{\rm Ar} \simeq 6.63 \times 10^{-26} \,{\rm kg}$.
Taking the reference (and hence, the wall) temperature to be 
$\widetilde{T}_{\rm ref} = \widetilde{T}_w = 273\, {\rm K}$,
the reference velocity \eqref{eq:cref} is 
$\widetilde{c}_{\rm ref} \simeq 238.35\,{\rm m}/{\rm s}$
and the sound speed is $\widetilde{c}_s \simeq 307.71\,{\rm m}/{\rm s}$ at 
$\widetilde{T} = \widetilde{T}_w$.
The average particle number density was taken to be 
$\widetilde{n} = 1.4\times 10^{20} \,{\rm molecules} / {\rm m}^3$
and the mean free path is $\widetilde{\lambda} = 0.008833\, {\rm m}$. 
The Knudsen number ${\rm Kn}$ is thus controlled by varying the 
domain size, such that the reference length and reference time depend on ${\rm Kn}$, 
as shown in Table~\ref{tab:ref}. 

In the Maxwell molecules model, the viscosity coefficient has a linear temperature 
dependence:
\begin{equation}
 \widetilde{\mu} = \widetilde{\mu}_{\rm ref} \frac{\widetilde{T}}{\widetilde{T}_{\rm ref}},
\end{equation}
where $\widetilde{\mu}_{\rm ref}$ is the viscosity at the reference temperature.
This expression for the viscosity can be achieved within the single
relaxation time approximation by setting the
non-dimensionalized relaxation time to \cite{bird94}:
\begin{equation}
 \tau = \frac{\rm Kn}{n},\label{eq:mw_tau}
\end{equation}
This expression ensures that the viscosity obtained via the Chapman-Enskog expansion
is linear with respect to the temperature:
\begin{equation}
 \mu = {\rm Kn} \, T.
\end{equation}
The Prandtl number is fixed at ${\rm Pr} = 2/3$ using the Shakhov model, as 
discussed in Sec.~\ref{sec:shk}.

In this section, we compare the results obtained using 
our LB models and the 
DSMC results at the level of the profiles of the density $n$, pressure $P$, 
temperature $T$, viscous stress $\Pi_{ij} = T_{ij} - n T \delta_{ij}$, 
velocity $u_y$ and heat fluxes $q_x$ and $q_y$. We considered three batches of 
simulations, which are discussed below.
For all simulations, a grid with $S = 16$ nodes, stretched according to 
Eq.~\eqref{eq:stretch}
with $A = 0.98$, was employed. The time step was set to $\delta t = 5 
\times 10^{-4}$. 
The model used was ${\rm HHLB}(6;7) \times {\rm HLB}(6;7)$ for 
all simulations, except at ${\rm Kn} = 1$. Since at ${\rm Kn} = 1$, the 
flow enters the transition regime, the quadrature order had to be increased to 
$Q_x = 11$ and the simulations in this regime were performed with the 
model ${\rm HHLB}(6; 11) \times {\rm HLB}(6;7)$. We note that 
increasing the expansion order of $g_x$ with respect to the half-range 
Hermite polynomials from $N_x = 6$ to higher values does not have any 
visible influence on the results.

In the first batch of simulations, the relative wall velocity difference 
is fixed at $2\widetilde{u}_w = 300\ {\rm m} / {\rm s}$ ($u_w \simeq 0.63$) and 
${\rm Kn}$ is varied from $0.01$ up to $1$. Our simulation results 
are shown in Fig.~\ref{fig:mw-u063} alongside the DSMC results. 
An excellent agreement can be seen for the velocity (a),
the heat fluxes [(b) and (c)],
the density (d) and
the non-diagonal component $\Pi_{xy}$ of the viscous stress tensor
 (i). The temperature, the pressure 
and the diagonal components of the stress tensor present visible
deviations when ${\rm Kn} \gtrsim 0.25$.

The second and third simulation batches are performed at ${\rm Kn} = 0.1$ 
and ${\rm Kn} = 0.5$, respectively, for values of the relative wall velocity difference
$2\widetilde{u}_w$ between $200\ {\rm m} / {\rm s}$ (${\rm Ma} \simeq 0.65$) 
and $1000\ {\rm m} / {\rm s}$ (${\rm Ma} \simeq 3.25$),
corresponding to $u_w \simeq 0.42$ and $u_w \simeq 2.1$, respectively.
The simulation results are shown in Figs.~\ref{fig:mw-kn01} and \ref{fig:mw-kn05}. In these figures,
a very good agreement can be seen between the LB and DSMC results.
It is interesting to note that there is some disagreement in the 
results for $\Pi_{ij}$, $T$ and $P$ at high wall velocities ($u_w \gtrsim 1.68$),
 even  at ${\rm Kn} = 0.1$. This disagreement seems to indicate that the 
 relaxation time model 
becomes inaccurate at high shearing rates. 

The comparisons presented in this section validate the
LB models with mixed Gauss-Hermite quadratures, for a wide range of the Knudsen 
number, as well as of the plate velocities. The 
simulations were performed using a discretization 
of the velocity space employing $2Q_x Q_y = 98$ distinct vectors 
($Q_x = Q_y = 7$) for 
${\rm Kn} < 1$ and $154$ distinct
velocities ($Q_x = 11$, $Q_y= 7$) at ${\rm Kn} = 1$. This makes our 
proposed method highly efficient for the study of channel flows in the 
rarefied regime.

\section{Comparison with DSMC: Realistic potentials}\label{sec:sharipov}

\begin{figure}
\begin{center}
\begin{tabular}{c}
 \includegraphics[width=0.95\columnwidth]{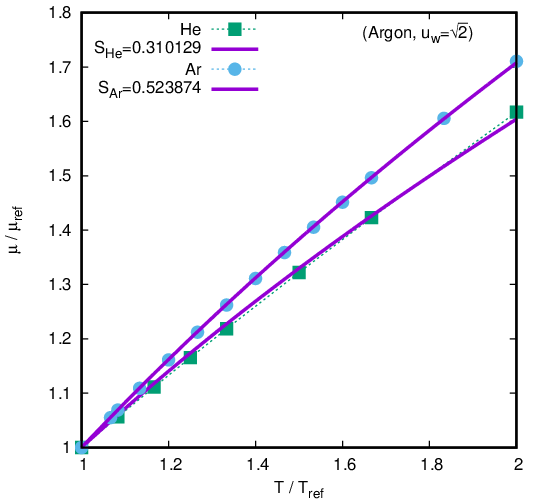}
\end{tabular}
\caption{Comparison between the tabulated values for the viscosity of He and Ar 
from Refs.~\cite{cencek12} and \cite{vogel10} (dotted lines and symbols) and 
the fit obtained using the Sutherland model \eqref{eq:sutherland_mu}. 
The reference values are 
$\widetilde{\mu}_{\rm ref}^{\rm He} \simeq 19.91\ {\rm \mu Pa} \cdot {\rm s}$ and 
$\widetilde{\mu}_{\rm ref}^{\rm Ar} \simeq 22.67\ {\rm \mu Pa} \cdot {\rm s}$, 
which represent the viscosity values at $\widetilde{T}_{\rm ref} = 300\ {\rm K}$. 
The values of the Sutherland parameter
are $S_{\rm He} = 0.3101$ and $S_{\rm Ar} = 0.5239$ after non-dimensionalization.
}
\label{fig:sutherland}
\end{center}
\end{figure}

\begin{figure*}
\begin{center}
\begin{tabular}{ccc}
 \includegraphics[width=0.32\linewidth]{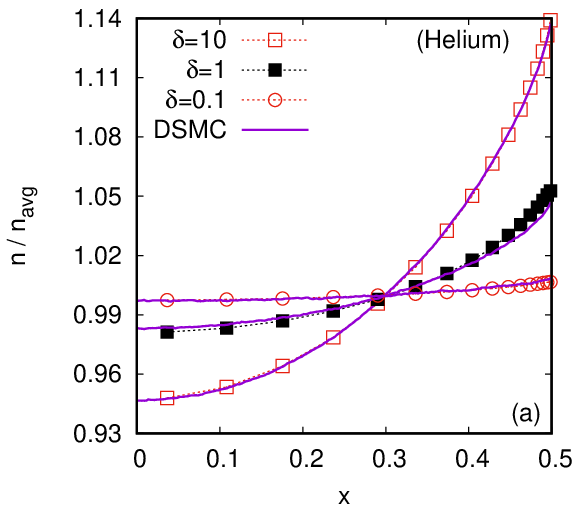} &
 \includegraphics[width=0.32\linewidth]{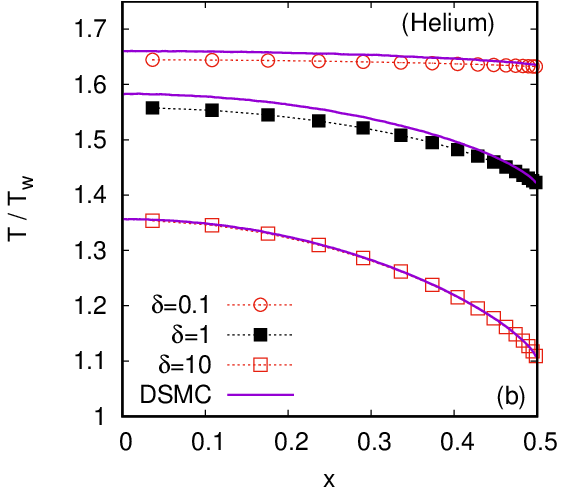} &
 \includegraphics[width=0.32\linewidth]{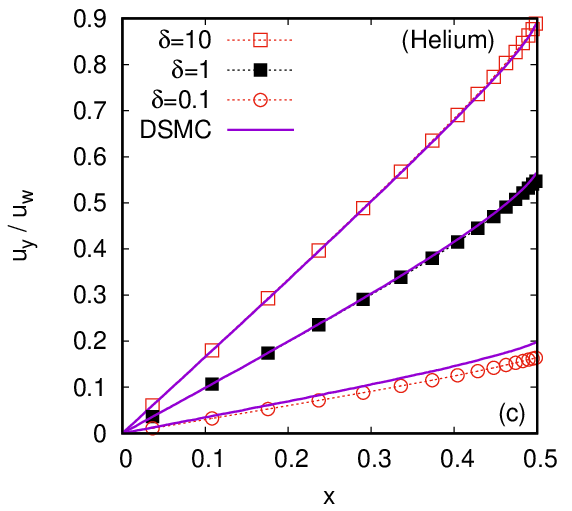} \\
 \includegraphics[width=0.32\linewidth]{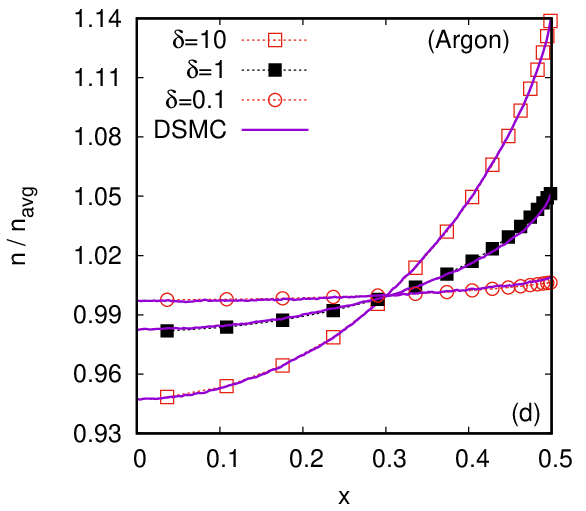} &
 \includegraphics[width=0.32\linewidth]{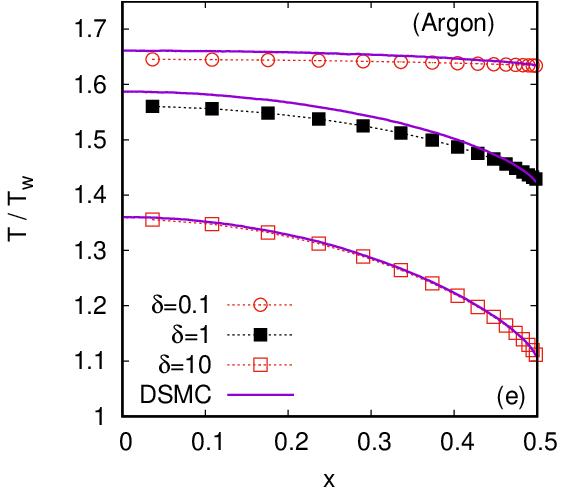} &
 \includegraphics[width=0.32\linewidth]{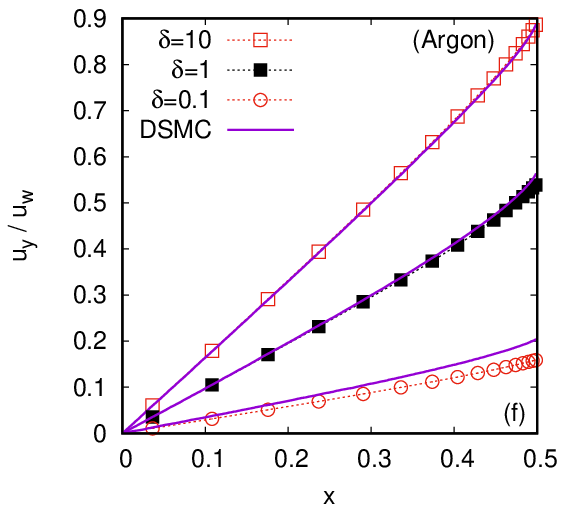}
\end{tabular}
\caption{Comparison of the LB results (dotted lines and points) and DSMC results 
from Ref.~\cite{sharipov13} (lines) for Helium (top) and 
Argon (bottom) molecules modeled using ab-initio potentials, 
at various values of the rarefaction parameter $\delta$.
The wall velocity is $u_w = \sqrt{2}$ and the relaxation time is implemented 
using the Sutherland model \eqref{eq:sutherland_tau}.
(left) Density $n$; (middle) Temperature $T$; (right) velocity $u_y$.
}
\label{fig:sharipov}
\end{center}
\end{figure*}

\begin{figure}
\begin{center}
\begin{tabular}{c}
 \includegraphics[width=0.95\columnwidth]{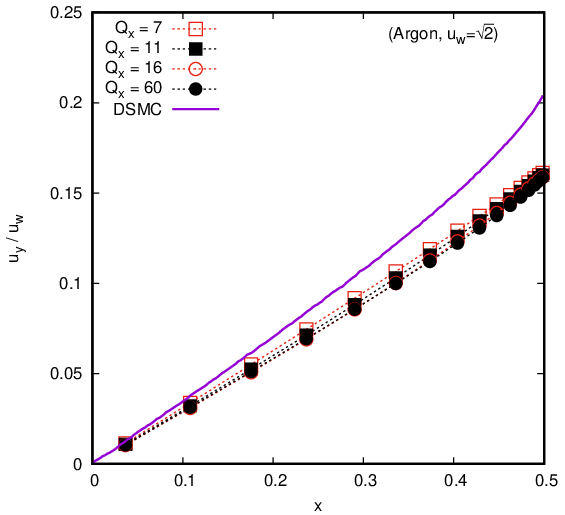}
\end{tabular}
\caption{Convergence process with respect to the quadrature order $Q_x$ 
for the Sutherland model for Argon molecules, compared with 
the DSMC results reported in Ref.~\cite{sharipov13} at $\delta = 0.1$. It 
can be seen that increasing $Q_x$ causes the LB results to depart from the 
DSMC profile.
}
\label{fig:sharipov_conv}
\end{center}
\end{figure}

\begin{figure*}
\begin{center}
\begin{tabular}{c}
 \includegraphics[width=0.33\linewidth]{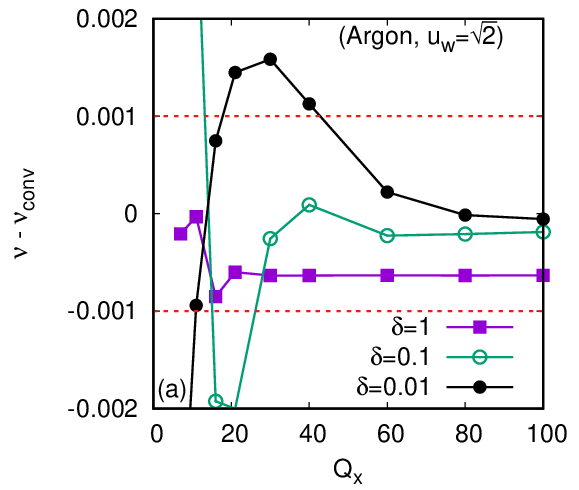}
 \includegraphics[width=0.33\linewidth]{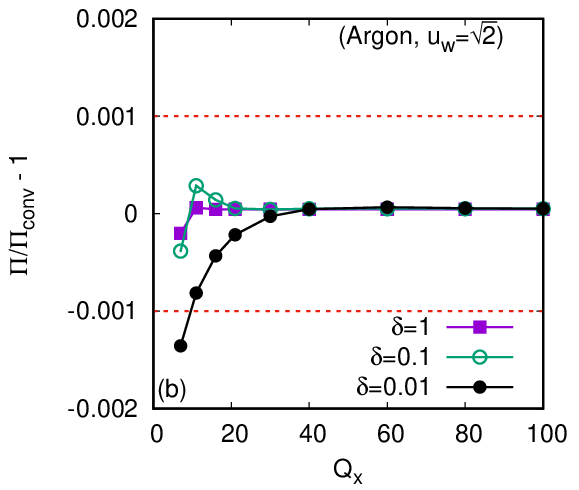}
 \includegraphics[width=0.33\linewidth]{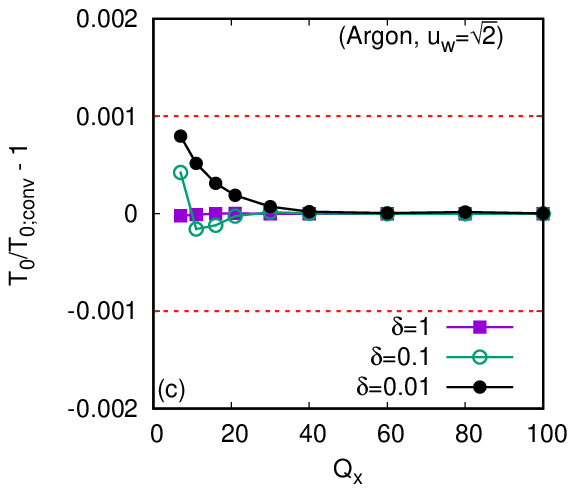}
\end{tabular}
\caption{Departure of the LB results for $\nu$, $\Pi$, and $T_0$ 
obtained at $u_w = \sqrt{2}$ for the Argon gas
with the models ${\rm HHLB}(6; Q_x) \times {\rm HLB}(6;7)$
on a grid with $S = 16$ nodes ($A = 0.98$),
with respect to the convergence values $\{\nu_{\rm conv}, \Pi_{\rm conv},
T_{0;{\rm conv}}\}$ obtained using the model ${\rm HHLB}(6; 100) \times {\rm HLB}(6;7)$
on a grid with $S = 32$ nodes ($A =0.98$), for $\delta \in \{0.1, 1, 10\}$.
(a) Absolute departure $\nu(Q_x) - \nu_{\rm conv}$; 
(b) Relative departure $\Pi(Q_x) /\Pi_{\rm conv} - 1$; 
(c) Relative departure $T_0(Q_x) / T_{0;{\rm conv}} - 1$.
}
\label{fig:sharipov_err}
\end{center}
\end{figure*}

\begin{table}
\begin{tabular}{|l|r|r|}
\hline
 $\delta$ & $Q_x$ &$\delta t$\\\hline
 $\ge 1$ & $7$ & $5 \times 10^{-4}$ \\
 $0.1$ & $30$ & $2.5 \times 10^{-4}$ \\
 $0.01$ & $100$ & $2.5 \times 10^{-4}$ \\\hline
\end{tabular}
\caption{Simulation parameters for the results shown in 
Fig.~\ref{fig:sharipov} and in Tables~\ref{tab:sharipov_gradu}, 
\ref{tab:sharipov_shear} and \ref{tab:sharipov_temp}. The simulations 
were conducted on a grid with $S = 16$ nodes, stretched according 
to Eq.~\eqref{eq:stretch} with $A = 0.98$. The expansion order 
of the equilibrium distribution with respect to the half-range Hermite 
polynomials was always kept at $N_x = 6$.
}
\label{tab:sharipov}
\end{table}

We now consider the validation of our LB models
in the case of the Couette flow of two noble gases, namely
Helium (He) and Argon (Ar).
The transport coefficients for these gases were measured experimentally 
and the experimental data can be found in Ref.~\cite{kestin84}.
More recently, these transport coefficients were computed 
using ab initio potentials and the results were reported in Refs.~\cite{cencek12}
and \cite{vogel10} for dilute Helium and Argon gases.
In this subsection, we compare our LB
simulation results with the results 
reported in Ref.~\cite{sharipov13}, calculated using the direct
simulation Monte Carlo (DSMC) method based on ab initio potentials over 
a wide range of the gas rarefaction parameter $\delta$. 
The results reported in Ref.~\cite{sharipov13}
concern mixtures of He and Ar, including the limiting cases of pure He and pure Ar. 
The treatment of gas mixtures requires more elaborate models, such as the 
McCormack model,
which was introduced in Ref.~\cite{mccormack73} for the linearized Boltzmann 
equation. Outside the linear regime, a relaxation time model for isothermal
 binary fluids was proposed in Ref.~\cite{sofonea01}.
To the best of our knowledge, there is no established relaxation time model 
which can be used to simulate non-isothermal gas mixtures. 
For simplicity, in this subsection we only consider the case of pure
monatomic gases (He or Ar), which can be easily
treated in the framework of the Shakhov model.

The Maxwell molecules model can be considered as a particular case of 
interaction model, for which the dynamic viscosity $\widetilde{\mu}$ varies 
with temperature according to:
\begin{equation}
 \widetilde{\mu} = \widetilde{\mu}_{\rm ref} 
 \left(\frac{\widetilde{T}}{\widetilde{T}_{\rm ref}}\right)^\omega,
\end{equation}
where $\widetilde{\mu}_{\rm ref}$ is the value of the viscosity at 
the reference temperature $\widetilde{T}_{\rm ref}$ 
and the viscosity index $\omega$ takes the value $\omega_{\rm Maxwell} = 1$ 
in the case of Maxwell molecules. In a more general formulation, the 
variable hard sphere (VHS) model gives rise to values of the viscosity 
index of the form $\omega = \frac{1}{2} + \nu$, where 
$\nu$ is a constant controlling the dependence of the collisional cross section on the 
relative speed of the interacting particles \cite{bird94}. The particular case of 
the hard sphere (HS) model is recovered by setting $\nu = 0$.
The VHS paradigm works remarkably well for quasi-isothermal flows. 
However, when the temperature variations 
in the flow are large, the data tabulated in Refs.~\cite{cencek12,vogel10} for the transport 
coefficients of various gases indicate that the viscosity index $\omega$ is a slowly-varying 
function of the temperature \cite{sharipov13}. 

In order to obtain reasonable agreement between the collisional model 
employed in DSMC and the
realistic data available for the transport coefficients, the generalized hard 
sphere (GHS)
model was introduced in Ref.~\cite{hassan93}. This model can reproduce with remarkable 
accuracy the experimental data in Ref.~\cite{kestin84} 
to temperatures as high as $1500\ {\rm K}$, where the 
deviation from the experimental data is about $5\%$ \cite{macrossan03}. 

In this section, we consider the viscosity law which emerges from the model 
proposed by Sutherland \cite{sutherland1893}.
In this model, the interaction potential is
considered to be the superposition between 
a hard-sphere-like infinite potential barrier around the 
repulsive core, followed by an attractive tail. This potential can be used to 
derive the scattering cross-sections of two-particle collisions and 
the value of the viscosity coefficient emerging from this model is given by 
\cite{sutherland1893}:
\begin{equation}
 \widetilde{\mu} = \widetilde{\mu}_{\rm ref} \left(
 \frac{\widetilde{T}}{\widetilde{T}_{\rm ref}}\right)^{1/2} 
 \frac{1 + \widetilde{S} / \widetilde{T}_{\rm ref}}
 {1 + \widetilde{S} / \widetilde{T}},\label{eq:sutherland_mu}
\end{equation}
where $\widetilde{\mu}_{\rm ref}$ is the
fluid viscosity at the temperature
$\widetilde{T} = \widetilde{T}_{\rm ref}$ and 
$\widetilde{S}$ is Sutherland's constant, which has the dimension of temperature. 
According to Ref.~\cite{macrossan03},
the Sutherland model can be fitted in order to reproduce with reasonable 
accuracy the experimental data reported in Ref.~\cite{kestin84}. 

In order to validate our simulation results against the results reported in Ref.~\cite{sharipov13},
we take the reference temperature to be equal to the wall temperature, which was set to 
$\widetilde{T}_{\rm ref} = 300\ {\rm K}$ therein. 
At this temperature, the values of the viscosity for ${\rm He}$ and ${\rm Ar}$ 
reported in Ref.~\cite{kestin84} are 
$\widetilde{\mu}_{\rm ref}^{\rm He} \simeq 20.04\ {\rm \mu Pa} \cdot {\rm s}$ and 
$\widetilde{\mu}_{\rm ref}^{\rm Ar} \simeq 22.83\ {\rm \mu Pa} \cdot {\rm s}$.
It can be seen that the viscosity of He given above does not
coincide with the value obtained via the ab initio formulation, namely 
$\widetilde{\mu}_{\rm ref; ab\ initio}^{\rm He} \simeq 19.91\ {\rm \mu Pa} \cdot {\rm s}$ 
\cite{cencek12,sharipov13}. Thus, in deriving the value of the Sutherland constant 
$\widetilde{S}$, we consider the values of the transport coefficients 
obtained in the framework of the ab initio calculations, 
as presented in Refs.~\cite{cencek12} and \cite{vogel10} for 
Helium and Argon, respectively.

Since the maximum value of the temperature 
attained in Ref.~\cite{sharipov13} is 
$\lesssim 2 \widetilde{T}_{\rm ref} = 600\ {\rm K}$,
we seek the values of $\widetilde{S}$ in Eq.~\eqref{eq:sutherland_mu}
which best reproduce the reference data over the temperature range 
$300\ {\rm K}$ - $600\ {\rm K}$. A non-linear fit gives 
the following values for $\widetilde{S}$:
\begin{align}
 \widetilde{S}_{\rm He} \simeq& 93.0387 \pm 3.159\ {\rm K}, \nonumber\\
 \widetilde{S}_{\rm Ar} \simeq& 157.1621 \pm 0.4047\ {\rm K}.
\end{align}
With the above choice of parameters, the maximum relative 
deviation of the viscosity from the tabulated data is less 
than $0.8 \%$ and $0.2 \%$ for ${\rm He}$ and ${\rm Ar}$, 
respectively. The result of the fit is shown in Fig.~\ref{fig:sutherland}.

The Sutherland model can be implemented by setting the
non-dimensional relaxation time $\tau$ within the Shakhov model to:
\begin{equation}
 \tau = \frac{1}{n T^{1/2} \delta \sqrt{2}} \frac{1 + S}{1 + S / T},
 \label{eq:sutherland_tau}
\end{equation}
where $\delta$ is the rarefaction parameter employed in Ref.~\cite{sharipov13},
while $S_{\rm He} \simeq 0.3101$ and $S_{\rm Ar} \simeq 0.5239$ after 
non-dimensionalization.

The profiles of $n$, $T$ and $u_y$ for $\delta = 0.1$, $1$ and $10$ at 
wall velocity difference $2u_w = 2\sqrt{2}$ (${\rm Ma} \simeq 2.2$) 
are shown in Fig.~\ref{fig:sharipov}. Good agreement 
can be observed in general, with the largest discrepancies occuring 
in the temperature profile for $\delta = 1$ and in the velocity 
profile at $\delta = 0.1$. For $\delta = 10$ and $1$, the simulations results
were obtained with the ${\rm HHLB}(6;7) \times {\rm HLB}(6;7)$ model
using $S = 16$ lattice nodes 
stretched according to Eq.~\eqref{eq:stretch} with $A = 0.98$ and 
the time step $\delta t = 5 \times 10^{-4}$. 
At $\delta = 0.1$, the flow is within the transition regime and the 
simulation results obtained with $Q_x = 7$ are no longer accurate within 
the S-model.
We thus obtained the results at $\delta = 0.1$ 
using the ${\rm HHLB}(6;30) \times {\rm HLB}(6; 7)$ model with $S = 16$, 
$A = 0.98$ and $\delta t = 2.5 \times 10^{-4}$ (increasing $N_x$ 
to values higher than $6$ did not make any visible differences to 
the results). The simulation parameters are summarized in Table~\ref{tab:sharipov}.
We note that increasing $Q_x$ did not bring the LB simulation results 
closer to the DSMC profiles, as indicated in Fig.~\ref{fig:sharipov_conv}.
This seems to indicate the fundamental limitation of the relaxation time approach,
which fails to provide accurate results far within the transition regime.

In order to validate our model with the results reported in Ref.~\cite{sharipov13}, we 
perform simulations at wall velocity differences $2u_w = 2\sqrt{2}$ (${\rm Ma} \simeq 2.2$)
and $2u_w = 0.2 \sqrt{2}$ (${\rm Ma} \simeq 0.22$), corresponding 
to $U = 2 v_0$ and $U = 0.2 v_0$ in Ref.~\cite{sharipov13}, for various 
values $\delta \in \{0.01, 0.1, 1, 10, 20, 40\}$ of the rarefaction parameter, 
covering the slip, transition and free molecular flow regimes.
The validation is performed at a quantitative level 
based on the
numerical results for the gradient $\nu$ of the velocity at the 
center of the channel (Table~\ref{tab:sharipov_gradu}),  the
value $\Pi$ of the shear stress (Table~\ref{tab:sharipov_shear}) and 
the value $T_0$ of the temperature measured in
the center of the channel
(Table~\ref{tab:sharipov_temp}). In all cases,
the LB results obtained using the above mentioned
values of the simulation parameters were compared with the 
results from Ref.~\cite{sharipov13}. The discrepancy is quantified by the 
relative error:
\begin{equation}
 \varepsilon(A) = \left|\frac{A_{\rm LB}}{ A_{\rm ref}} - 1\right|,
\end{equation}
where $A_{\rm LB}$ represents our simulation result and $A_{\rm ref}$ is the 
reference value from Ref.~\cite{sharipov13}. The comparison is performed 
for $A \in \{\nu, \Pi, T_0\}$, where the values of the
velocity gradient $\nu$, of the
temperature $T_0$  (calculated in the channel 
center), as well as of the quantity $\Pi$ (derived from 
the non-diagonal stress $T_{xy}$), are introduced below.

We begin by considering the dimensionless velocity gradient defined in 
Ref.~\cite{sharipov13}, which in our non-dimensionalization convention reads:
\begin{equation}
 \nu = \frac{L}{2u_w} \left.\frac{du_y}{dx}\right\rfloor_{x = 0}.
 \label{eq:vgrad}
\end{equation}
The derivative appearing above is computed using Eq.~\eqref{eq:ud0_extra}.
As can be seen in Table~\ref{tab:sharipov_gradu}, our results are in 
very good agreement with those reported in Ref.~\cite{sharipov13} for 
both He and Ar for $\delta \ge 1$. These results were obtained using the 
${\rm HHLB}(6;7) \times {\rm HLB}(6; 7)$ model and have the absolute error 
bounded by $\pm 0.001$. At $\delta = 0.1$ and $0.01$, in order to obtain LB results 
with the same $\pm 0.001$ absolute error, the quadrature order was raised up to 
$Q_x = 30$ and $100$, respectively. For the convenience of our readers, 
these simulation parameters are summarized in Table~\ref{tab:sharipov}.
The effect of increasing $Q_x$ at 
small values of $\delta$ on the absolute error $\nu(Q_x) - \nu_{\rm conv}$ 
computed with respect to the 
value $\nu_{\rm conv}$ obtained using the ${\rm HHLB}(6;100) \times {\rm HLB}(6; 7)$ 
model on a grid with $S = 32$ points ($A = 0.98$) is shown in 
Fig.~\ref{fig:sharipov_err}(a).
The necessity to increase $Q_x$ as $\delta$ is decreased was also demonstrated in 
Fig.~\ref{fig:slip_err}(b) in the context of the analysis of the linearized limit 
of the Boltzmann equation discussed in Sec.~\ref{sec:lowmach} (a decreasing value of 
$\delta$ corresponds to an increasing value of $k$ in Fig.~\ref{fig:slip_err}).

Next, we consider the non-dimensional
quantity $\Pi$, defined as \cite{sharipov13}:
\begin{equation}
 \Pi = -\frac{\widetilde{T}_{xy}}{\widetilde{n}_{\rm avg} \widetilde{u}_w 
 \sqrt{2 \widetilde{m} \widetilde{K}_B \widetilde{T}_w}}
 = -\frac{T_{xy}}{u_w \sqrt{2}},\label{eq:stress}
\end{equation}
where $\widetilde{n}_{\rm avg}$ represents the average density inside the channel.
Although $\Pi$ should be constant in
the stationary state of the Couette flow, 
small fluctuations of this quantity
are always present across the channel.
For this reason, $\Pi$ is computed using 
the average value of $T_{xy}$, obtained according to Eq.~\eqref{eq:Txy}.
Table~\ref{tab:sharipov_shear} summarizes our results for 
$u_w \in \{\sqrt{2}, 0.1\sqrt{2}\}$. It can be seen that the relative error 
between the LB and DSMC results is less than $2\%$ for all tested values of the 
parameters. For consistency with the results reported for $\nu$ in 
Table~\ref{tab:sharipov_gradu}, the models summarized in Table~\ref{tab:sharipov}
were employed. However, we note that the values of $\Pi$ obtained 
using the ${\rm HHLB}(6;7) \times {\rm HLB}(6;7)$ model are within 
$0.2\%$ error with respect to the LB values inscribed in the table, as 
can be seen from Fig.~\ref{fig:sharipov_err}(b). This observation is 
consistent with the results presented in Fig.~\ref{fig:slip_err}(d) 
for the error of $T_{xy}$ as compared with the solution of the 
linearized Boltzmann equation.

Finally, the temperature $T_0$ in the channel center is obtained using the 
following formula:
\begin{multline}
 T_0(x = 0) = \frac{x_2^2 x_3^2 T_1}{(x_1^2 - x_2^2)(x_1^2-x_3^2)} \\
 + \frac{x_3^2 x_1^2 T_2}{(x_2^2 - x_1^2)(x_2^2-x_3^2)} 
 + \frac{x_1^2 x_2^2 T_3}{(x_3^2 - x_1^2)(x_3^2-x_2^2)},
 \label{eq:T0}
\end{multline}
which is sixth order accurate with respect to the spacing $\delta \eta$ for 
even functions of $x$. In the above, $T_s$ corresponds to the temperature 
$T(x_s)$ at point $x_s$, which is given by Eq.~\eqref{eq:etas}.
Table~\ref{tab:sharipov_temp} shows a comparison between the LB and the DSMC 
results for the temperature $T_0$ in the center of the channel, obtained for $u_w = \sqrt{2}$
and various values of $\delta$. As was the case for $\Pi$, the results shown in the 
table were obtained using the models summarized in Table~\ref{tab:sharipov}, however,
the results obtained using the ${\rm HHLB}(6;7) \times {\rm HLB}(6;7)$ model 
are within less than $0.1\%$ relative error with respect to the LB results obtained 
with the model ${\rm HHLB}(6;100) \times {\rm HLB}(6;7)$ on a finer grid ($S = 32$, 
$A =0.98$), as can be seen in Fig.~\ref{fig:sharipov_err}(c). 

We end this section with a comment on the accuracy of the simulation results presented 
herein. The results reported in Fig.~\ref{fig:sharipov} and in 
Tables~\ref{tab:sharipov_gradu}--\ref{tab:sharipov_temp} were obtained on a grid with 
$S = 16$ nodes, stretched according to Eq.~\eqref{eq:etas} with $A= 0.98$. 
The time step was set to $\delta t = 5 \times 10^{-4}$ for $Q_x = 7$ and 
$\delta = 2.5 \times 10^{-4}$ for $Q_x = 30$ and $100$, as summarized in Table~\ref{tab:sharipov}.
The results shown in Table \ref{tab:sharipov_gradu} for $\nu$ have a maximum 
absolute error of $0.001$, 
while the results for $\Pi$ and $T_0$ shown in Tables \ref{tab:sharipov_shear} 
and \ref{tab:sharipov_temp} have relative errors of less than $0.1\%$.
We checked that the effects of halving $\delta t$ or doubling $S$ 
were within these error bounds. 

A comparison of the LB and the DSMC data for $\Pi$ and $T_0$ 
shown in Tables~\ref{tab:sharipov_shear} and \ref{tab:sharipov_temp}
together with the convergence analysis shown in Fig.~\ref{fig:sharipov_err},
panels (b) and (c), reveals that the Shakhov model can provide accurate estimates 
of the DSMC results (relative errors below $2\%$) with a very modest cost, since 
the results obtained 
with the ${\rm HHLB}(6;7) \times {\rm HLB}(6;7)$ on a grid with $16$ nodes 
have less than $0.2\%$ error with respect to the results obtained using the 
${\rm HHLB}(6; 100)\times {\rm HLB}(6;7)$ model on a grid with $32$ nodes.
In terms of absolute error, the values of the
velocity gradient $\nu$, obtained using our LB 
models are within less than $5\%$ of the DSMC results. However, because
 the value of $\nu$ 
decreases as $\delta$ is decreased, the relative error becomes very large 
when $\delta \lesssim 0.1$. Furthermore, accurate estimates of $\nu$ 
using our LB models require the quadrature order $Q_x$ to be increased to 
large values as $\delta$ is decreased ($Q_x = 100$ was employed at $\delta = 0.01$).

\begin{table}
\begin{tabular}{|c|ccc|ccc|}
\hline 
 & \multicolumn{3}{c|}{$\nu$ (Argon)} & \multicolumn{3}{c}{$\nu$ (Helium)} \\
 $\delta$ & LB & Ref.~\cite{sharipov13} & $\varepsilon(\nu_{\rm Ar})$ &
 LB & Ref.~\cite{sharipov13} & $\varepsilon(\nu_{\rm He})$ \\\hline
 $0.01$ & $0.027$ & $0.048$ & $43.8\%$ & $0.028$ & $0.041$ & $31.7\%$ \\
 $0.1$ & $0.145$ & $0.173$ & $16.2\%$ & $0.149$ & $0.172$ & $13.40\%$ \\
 $1$ & $0.484$ & $0.486$ & $0.42\%$ & $0.493$ & $0.494$ & $0.20\%$ \\
 $10$ & $0.824$ & $0.819$ & $0.62\%$ & $0.831$ & $0.826$ & $0.61\%$ \\
 $20$ & $0.874$ & $0.873$ & $0.12\%$ & $0.880$ & $0.880$ & $0.0\%$ \\
 $40$ & $0.905$ & $0.904$ & $0.12\%$ & $0.911$ & $0.914$ & $0.33\%$ \\\hline
\end{tabular}
\caption{Comparison of the results obtained for the velocity gradient $\nu$ \eqref{eq:vgrad}
using the LB model employed in this paper and the DSMC data reported in Ref.~\cite{sharipov13},
for $u_w = \sqrt{2}$ and various values of $\delta$.
\label{tab:sharipov_gradu}}
\end{table}

\begin{table}
\begin{tabular}{|c|ccc|ccc|}
\hline
 & \multicolumn{3}{c|}{$\Pi$ (Argon)} & \multicolumn{3}{c|}{$\Pi$ (Helium)} \\
 $\delta$ & LB & Ref.~\cite{sharipov13} & $\varepsilon(\Pi_{\rm Ar})$ &
 LB & Ref.~\cite{sharipov13} & $\varepsilon(\Pi_{\rm He})$ \\\hline
 \multicolumn{7}{|c|}{$u_w = \sqrt{2}$}\\\hline
 $0.01$ & $0.5619$ & $0.5612$ & $0.12\%$ & $0.5618$ & $0.5615$ & $0.05\%$ \\
 $0.1$ & $0.5358$ & $0.5319$ & $0.73\%$ & $0.5344$ & $0.5315$ & $0.55\%$ \\
 $1$ & $0.3720$ & $0.3663$ & $1.56\%$ & $0.3671$ & $0.3616$ & $1.52\%$ \\
 $10$ & $0.09787$ & $0.09777$ & $0.10\%$ & $0.09585$ & $0.09551$ & $0.36\%$\\
 $20$ & $0.05328$ & $0.05316$ & $0.23\%$ & $0.05223$ & $0.05191$ & $0.62\%$\\
 $40$ & $0.02769$ & $0.02766$ & $0.11\%$ & $0.02718$ & $0.02704$ & $0.52\%$ \\\hline
 \multicolumn{7}{|c|}{$u_w = 0.1 \sqrt{2}$}\\\hline
 $0.01$ & $0.5594$ & $0.5575$ & $0.34\%$ & $0.5594$ & $0.5585$ & $0.16\%$ \\
 $0.1$ & $0.5225$ & $0.5167$ & $1.12\%$ & $0.5225$ & $0.5191$ & $0.65\%$ \\
 $1$ & $0.3392$ & $0.3365$ & $0.81\%$ & $0.3392$ & $0.3382$ & $0.30\%$ \\
 $10$ & $0.08324$ & $0.08320$ & $0.05\%$ & $0.08322$ & $0.08324$ & $0.03\%$ \\
 $20$ & $0.04546$ & $0.04531$ & $0.34\%$ & $0.04545$ & $0.04540$ & $0.12\%$ \\
 $40$ & $0.02383$ & $0.02381$ & $0.09\%$ & $0.02382$ & $0.02383$ & $0.04\%$ \\\hline
\end{tabular}
\caption{Comparison of the results obtained for the shear stress $\Pi$ \eqref{eq:stress}
using the LB model employed in this paper and the DSMC data reported in Ref.~\cite{sharipov13},
for $u_w \in \{\sqrt{2},0.1\sqrt{2}\}$ and various values of $\delta$. \label{tab:sharipov_shear}
}
\end{table}

\begin{table}
\begin{tabular}{|c|ccc|ccc|}
\hline
 & \multicolumn{3}{c|}{$T_0$ (Argon)} & \multicolumn{3}{c}{$T_0$ (Helium)} \\
 $\delta$ & LB & Ref.~\cite{sharipov13} & $\varepsilon(T_0^{\rm Ar})$ &
 LB & Ref.~\cite{sharipov13} & $\varepsilon(T_0^{\rm He})$ \\\hline
 $0.01$ & $1.663$ & $1.667$ & $0.24\%$ & $1.663$ & $1.665$ & $0.12\%$ \\
 $0.1$ & $1.646$ & $1.661$ & $0.91\%$ & $1.645$ & $1.660$ & $0.91\%$ \\
 $1$ & $1.561$ & $1.587$ & $1.64\%$ & $1.558$ & $1.583$ & $1.58\%$ \\
 $10$ & $1.357$ & $1.360$ & $0.22\%$ & $1.355$ & $1.356$ & $0.08\%$\\
 $20$ & $1.315$ & $1.316$ & $0.08\%$ & $1.314$ & $1.313$ & $0.08\%$ \\
 $40$ & $1.291$ & $1.291$ & $0\%$ & $1.291$ & $1.289$ & $0.16\%$\\\hline
\end{tabular}
\caption{Comparison of the results obtained for the temperature $T_0$ at the center 
of the channel using the LB model employed in this paper and the DSMC data 
reported in Ref.~\cite{sharipov13}
for $u_w = \sqrt{2}$ and various values of $\delta$.
\label{tab:sharipov_temp}}
\end{table}

\section{Conclusion}\label{sec:conc}

In this paper, we studied the 3D Couette flow using the mixed quadrature 
lattice Boltzmann models introduced in Ref.~\cite{ambrus16jcp}, which employ
the half-range Gauss-Hermite quadrature of order $Q_x$ 
on the axis perpendicular to the walls (the $x$ axis)
and the full-range Gauss-Hermite quadrature of order $Q_y$ 
on the axis parallel to the flow direction (the $y$ axis). 
The third degree of freedom in the momentum 
space was removed through 
the analytic integration of the Boltzmann equation and the subsequent analysis 
was performed using reduced distribution functions.

We first validated our LB models in the low Mach number regime 
by comparing our simulation results with the benchmark results obtained
in Ref.~\cite{jiang16} through a semi-analytic procedure applied to the 
linearized Boltzmann-BGK equation. To ensure that our simulations remained in 
the linearized regime, the wall velocity was set to a small value ($u_w = 10^{-5}$).
The validation was performed at the level of the velocity at the wall $u_y(L/2)$, 
the derivative of the velocity at the center of the channel $u'_y(0)$,
the non-diagonal component $T_{xy}$ of the 
stress tensor, and the half-channel mass flow rate $\dot{m}$.
By employing a convergence test, we concluded that the minimum 
quadrature order required in order
to achieve a given accuracy ($1\%$ error tolerance) must be increased 
as the Knudsen number $k$ is increased. 
Setting $Q_x= 4$ ensures that the 
relative errors in $T_{xy}$ are less than $1\%$ up to $k = 100$. 
For the remaining three quantities, the relative errors are below $1\%$ 
up to $k \lesssim 0.1$. Setting $Q_x = 7$ preserves the $1\%$ error 
threshold up to $k \simeq 1$, while the relative error in $T_{xy}$ is decreased 
below $0.1\%$ up to $k = 100$. Between $1 \lesssim k \lesssim 100$, 
the quadrature order has to be increased dramatically in order to ensure 
that the relative errors in $u_y(L/2)$, $u'_y(0)$ and $\dot{m}$ remain 
below $1\%$. This is partly due to the fact that the absolute values of these 
quantities decrease as $k$ is increased, such that maintaining a $1\%$ relative 
error entails an effective increase of the simulation accuracy. These convergence 
tests were performed employing a grid with $S = 16$ points spanning half of the 
flow channel, stretched towards the bounding wall. When comparing the nonlinear 
part of the velocity profile with the data reported in Ref.~\cite{li15}, we found 
very good agreement after refining the grid to $S = 64$ points for $k = 0.03$ 
($Q_x = 4$ was sufficient here) and $S = 32$ points for $k = 0.1$ ($Q_x = 7$), 
$k = 1$ ($Q_x = 21$) and $k = 10$ ($Q_x = 80$).

Next, we compared the LB profiles of the macroscopic quantities 
(particle number density, velocity, pressure tensor and heat flux)
with the Direct Simulation Monte Carlo (DSMC) results for Maxwell molecules 
reported in Refs.~\cite{struchtrup07,struchtrup08,torrilhon08,taheri09,schuetze03}.
The LB profiles were obtained for ${\rm Kn} < 1$ with
the ${\rm HHLB}(6;7) \times {\rm HLB}(6;7)$ model, 
employing $Q_x = Q_y = 7$ and an expansion up to $N_x = N_y = 6$ of the 
equilibrium distribution with respect to the half-range (on the $x$ axis) 
and full range (on the $y$ axis) Hermite polynomials. 
At ${\rm Kn} = 1$, the quadrature order on the $x$ axis 
was raised to $Q_x = 11$.
We found that the DSMC results for Knudsen numbers between 
$0.01 \le {\rm Kn} \le 1$ and for wall velocities between 
$0.42 \le u_w \le 2.1$ could be reasonably well recovered 
by employing the Shakhov collision term.
We found deviations between our LB results and the DSMC data at 
${\rm Kn} \gtrsim 0.25$, as well as
when the wall velocity exceeded 
$u_w \gtrsim 1.68$.

Finally, we performed a comparison with the results for pure Helium and Argon 
obtained in Ref.~\cite{sharipov13} using an interaction model based on 
ab initio potentials. In order to match the ab initio transport coefficients,
we implemented the Sutherland model and obtained the Sutherland constant 
by fitting the analytic expression for the viscosity to the
tabulated data reported in 
Refs.~\cite{cencek12} and \cite{vogel10}. The relative errors of the 
viscosity obtained in the frame of the Sutherland model compared to the 
tabulated data are below $1\%$ for the temperature range 
relevant for the simulations considered in this paper 
($300\ {\rm K}$ -- $600\ {\rm K}$). 
At the level of the profiles for the density $n$, velocity $u_y$ and 
temperature $T$, the LB results are in very good agreement with the DSMC data 
when the rarefaction parameter $\delta$ satisfies $\delta > 1$. 
At $\delta \lesssim 1$, visible deviations occur in the profile 
of $T$. Moreover, the velocity profile also deviates from the DSMC data 
at $\delta = 0.1$. This observed discrepancy is 
due to the limitations of the relaxation time approximation of the collision 
integral, since increasing the quadrature order does not bring our FDLB 
results closer to the DSMC data. A quantitative analysis at the level 
of the temperature $T_0$ at the channel center and 
of the shear pressure $\Pi$ shows that the deviations of our FDLB models 
from the DSMC results are within a few percent. The velocity gradient $\nu$ 
at the channel center presents an increasing relative error as $\delta$ is 
decreased, which may also be due to the fact that $\nu$ decreases towards $0$ as 
$\delta$ is decreased. In terms of an absolute error, the LB results for 
$\nu$ still remain within a few percent of the DSMC data (for the Argon 
gas at $\delta = 0.01$ and $u_w= \sqrt{2}$, $\nu_{\rm DSMC} = 0.048$, 
while the FDLB result is $\nu_{\rm FDLB} = 0.027$). 

The analysis presented in this 
paper indicates that the solution of the S-model equation seems to be 
within a few percent of the DSMC results in the context of the Couette 
flow for velocities up to $2.1$ and for values of the 
rarefaction parameter $\delta$ down to $0.01$ (values of the Knudsen number 
${\rm Kn}$ up to $\simeq 70$). 

We finish this paper by noting that the ${\rm HHLB}(6;7) \times {\rm HLB}(6;7)$ 
model, which employs only $2Q_x Q_y = 98$ distinct velocities, can be used to 
obtain a very good estimate (within $1\%$ relative error) of the 
solution of the Shakhov model equation at the level of the 
temperature in the channel center and non-diagonal component of the stress 
tensor. We thus conclude that the use of half-range 
quadratures is an essential ingredient when 
considering the channel flow of rarefied gases.

\section*{Acknowledgements}

This work was supported by a grant
of the Romanian National Authority for Scientific Research,
CNCS-UEFISCDI, Project No. PN-II-ID-PCE-2011-3-0516.
The authors express their gratitude to Professor
Henning Struchtrup (Department of Mechanical Engineering,
University of Victoria, Canada) for kindly providing the DSMC 
results for Maxwell molecules used 
for the validation of the models introduced in this paper. The authors are 
grateful to Prof. Li-Shi Luo (Department of Mathematics and Statistics, Old
Dominion University, Norfolk, VA) for suggesting the validation 
of their results by considering the non-linear part of the 
velocity profile in the linearized regime. The authors are indebted to 
Prof. Felix Sharipov (Universidade Federal do Paran\'a, Curitiba, Brazil)
for suggesting the validation tests in the context of realistic, 
ab-initio potentials and for kindly providing DSMC results for comparison.

\appendix*

\section{Numerical scheme}\label{app:scheme}

The simulation results presented in this paper were obtained using an
explicit third order total variation diminishing (TVD) Runge-Kutta (RK-3) 
time marching procedure \cite{gottlieb98,henrick05,shu88,trangenstein07},
as described in Subsec.~\ref{app:scheme:time}.
In order to increase the
simulation efficiency, we follow Refs.~\cite{guo03,mei98jcph} and 
employ a grid stretching algorithm 
 to increase the grid resolution 
in the vicinity of the solid wall, as described in Subsec.~\ref{app:scheme:stretch}.
The fifth-order weighted essentially non-oscillatory (WENO-5) scheme 
\cite{gan11,jiang96} employed for the advection is presented in 
Subsec.~\ref{app:scheme:adv}. 
Finally, the implementation of the diffuse reflection and bounce-back boundary conditions 
is discussed in Subsec.~\ref{app:scheme:bc}.

\subsection{Time stepping}\label{app:scheme:time}

In order to implement the time stepping algorithm, it is convenient 
to cast the Boltzmann equation \eqref{eq:boltz} in the following form:
\begin{equation}
 \partial_t F = L[F], \quad 
 L[F] = -\frac{p_x}{m} \partial_x F - 
 \frac{1}{\tau} \left[F - F^{(\rm eq)}(1 + \mathbb{S}_F)\right], \label{eq:L_def}
\end{equation}
where $F \in \{\phi, \chi\}$ is any of the two reduced distributions 
introduced in Sec.~\ref{sec:lb}, while $\mathbb{S}_F$ is given 
in Eq.~\eqref{eq:sred_def}.

Let us consider a discretization of the time coordinate using equal 
time steps $\delta t$, such that $t_\ell = \ell\, \delta t$.
The distribution functions at time step $\ell$ can be written 
as $F_{\ell} \equiv F(t_{\ell})$. The value of 
$F_{\ell + 1}$ can be obtained using the third-order Runge-Kutta 
TVD method introduced in Ref.~\cite{shu88}, using two intermediate steps,
as follows:
\begin{align}
 F_\ell^{(1)} =& F_\ell + \delta t \, L[F_\ell], \nonumber\\
 F_\ell^{(2)} =& \frac{3}{4} F_\ell + \frac{1}{4} F_\ell^{(1)} + 
 \frac{1}{4} \delta t\, L[F_\ell^{(1)}],\nonumber\\
 F_{\ell + 1} =& \frac{1}{3} F_\ell + \frac{2}{3} F_\ell^{(2)} + 
 \frac{2}{3} \delta t\, L[F_\ell^{(2)}]. \label{eq:rk3}
\end{align}
The Butcher tableau \cite{butcher08} for the above scheme is summarized 
in Table~\ref{tab:rk3}.

\begin{table}
\begin{center}
\begin{tabular}{r|rrr}
0 & & & \\
1 & 1 & & \\
1/2 & 1/4 & 1/4 & \\\hline
& 1/6 & 1/6 & 2/3
\end{tabular}
\end{center}
\caption{Butcher tableau for the third-order Runge-Kutta 
integration summarized in Eq.~\eqref{eq:rk3}.\label{tab:rk3}}
\end{table}

\subsection{Grid stretching}\label{app:scheme:stretch}

As highlighted in Refs.~\cite{guo03,mei98jcph}, a finer mesh is needed 
in the vicinity of solid boundaries as compared to the bulk regions of 
the flow in order to capture the Knudsen layer effects. This can be achieved 
by performing a standard grid-stretching procedure and in this paper,
we follow Ref.~\cite{busuioc17} and characterize the refined 
mesh using the non-dimensional parameter $\eta$ as follows:
\begin{equation}
 x(\eta) = \frac{L}{2A} \tanh{\eta},\label{eq:stretch}
\end{equation}
where $0 \le \eta \le {\rm arctanh}\, A$ and $0 < A < 1$ controls the 
stretching such that when $A \rightarrow 0$, the grid becomes equidistant with 
respect to $x$, while as $A \rightarrow 1$, the grid points accumulate towards the 
right boundary. 

The flow domain is discretized using $S$ equidistant values of 
$\eta$, namely:
\begin{equation}
 \eta_s = \frac{1}{S}\left(s - \frac{1}{2}\right) {\rm arctanh}\, A, \qquad 
 x_s = \frac{L}{2A} \tanh{\eta_s},\label{eq:etas}
\end{equation}
where the points with $1 \le s \le S$ lie within the flow domain. 
The value $A = 0.98$ was employed for all simulations presented in this paper.

\subsection{Advection}\label{app:scheme:adv}

\begin{table}
\begin{center}
\begin{tabular}{r|rrr}
 & $\overline{\omega}_1$ & $\overline{\omega}_2$ & $\overline{\omega}_3$ \\\hline
$\sigma_1 = \sigma_2 = \sigma_3 = 0$ & $0.1$ & $0.6$ & $0.3$ \\\hline
$\sigma_2 = \sigma_3 = 0$ & $0$ & $2/3$ & $1/3$ \\
$\sigma_3 = \sigma_1 = 0$ & $1/4$ & $0$ & $3/4$ \\
$\sigma_1 = \sigma_2 = 0$ & $1/7$ & $6/7$ & $0$ \\\hline
$\sigma_1 = 0$ & $1$ & $0$ & $0$\\
$\sigma_2 = 0$ & $0$ & $1$ & $0$\\
$\sigma_3 = 0$ & $0$ & $0$ & $1$
\end{tabular}
\end{center}
\caption{The limiting values of $\overline{\omega}_q$ \eqref{eq:weno5_omega} 
when any combination of indicator of smoothness functions $\sigma_i$ 
have vanishing values.\label{tab:weno}}
\end{table}

The spatial derivative occuring in Eq.~\eqref{eq:boltz} can be 
approximated by considering an equidistant grid with respect to 
the $\eta$ coordinate \eqref{eq:etas}:
\begin{equation}
 \left(\frac{p_x}{m} \partial_x F\right)_s =
 \frac{\mathcal{F}_{s+1/2} - \mathcal{F}_{s-1/2}}{x_{s+1/2} - x_{s - 1/2}}.
 \label{eq:flux}
\end{equation}
The flux $\mathcal{F}_{s+1/2}$ corresponds to the interface between 
the cells centered on $\eta_s$ and $\eta_{s+1}$, while the coordinates
$x_{s\pm 1/2}$ of these interfaces are obtained by substituting 
$\eta = \eta_s \pm \delta \eta / 2$ in Eq.~\eqref{eq:etas}.
The fluxes are computed using the WENO-5 algorithm 
\cite{gan11,blaga17prc,busuioc17,busuioc18}.
For the case when the advection velocity $p_x / m$ is positive, the flux 
is given by
\begin{equation}
\mathcal{F}_{s+1/2} = \overline{\omega}_1\mathcal{F}^1_{s+1/2} +
\overline{\omega}_2\mathcal{F}^2_{s+1/2} + \overline{\omega}_3\mathcal{F}^3_{s+1/2},
\label{eq:weno5_flux}
\end{equation}
where $\mathcal{F}^q_{s + 1/2}$ ($q = 1,2,3$) are interpolating functions, which 
can be computed as follows:
\begin{align}
\mathcal{F}^1_{s+1/2} =& \frac{p_x}{m} \left(\frac{1}{3}F_{s-2} - \frac{7}{6} F_{s-1} + \frac{11}{6} F_s\right), \nonumber \\
\mathcal{F}^2_{s+1/2} =& \frac{p_x}{m} \left(-\frac{1}{6}F_{s-1} + \frac{5}{6} F_{s} + \frac{1}{3} F_{s+1}\right), \nonumber \\
\mathcal{F}^3_{s+1/2} =& \frac{p_x}{m} \left(\frac{1}{3}F_{s} + \frac{5}{6} F_{s+1} - \frac{1}{6} F_{s+2}\right).
\end{align}
The weighting factors $\overline{\omega}_q$ are given by:
\begin{equation}
\overline{\omega}_q = \frac{\widetilde{\omega}_q}{\widetilde{\omega}_1+\widetilde{\omega}_2+\widetilde{\omega}_3}, \qquad 
\widetilde{\omega}_q = \frac{\delta_q}{\sigma^2_q}.\label{eq:weno5_omega}
\end{equation}
where $\delta_q \in \{0.1, 0.6, 0.3\}$ are the ideal weights.
The indicators of smoothness $\sigma_q$ can be computed using:
\begin{align}
\sigma_1 =& \frac{13}{12} \left(F_{s-2} -2F_{s-1} + F_s \right)^2 
+ \frac{1}{4} \left( F_{s-2} - 4F_{s-1} + 3F_s \right)^2, 
\nonumber \\
\sigma_2 =& \frac{13}{12} \left(F_{s-1} -2F_{s} + F_{s+1} \right)^2 
+ \frac{1}{4} \left( F_{s-1} - F_{s+1} \right)^2,
\nonumber \\
\sigma_3 =& \frac{13}{12} \left(F_{s} -2F_{s+1} + F_{s+2} \right)^2 
+ \frac{1}{4} \left( 3F_{s} -4 F_{s+1} + F_{s+2} \right)^2.
\label{eq:weno_sigma}
\end{align}
It is customary in numerical algorithms to add a small quantity 
$\varepsilon \simeq 10^{-6}$ to $\sigma_q$ in order to avoid
division by zero. This operation can have side effects 
which depend on the magnitude of the advected quantity $F$, 
as discussed in Ref.~\cite{henrick05}. In order to avoid 
such side effects, $\overline{\omega}_q$ is computed directly 
from Table~\ref{tab:weno} in the limiting cases when one or more 
of the $\sigma_q$ functions vanish.

\subsection{Boundary conditions}\label{app:scheme:bc}

The Couette flow considered in this paper is symmetric with respect 
to the channel centerline, thus allowing the simulation domain to be
reduced to only the right half of the channel, such that $0 \le x \le L/2$.
The symmetry condition of the Couette flow is immediately achieved when 
bounce-back boundary conditions are implemented on the centerline at $x= 0$. 
The gas-wall interaction is modeled using diffuse reflection boundary 
conditions \cite{ambrus16jcp,ambrus16jocs,jcph09},
which are implemented at $x = L/2$.
In order to apply the fifth order WENO scheme described in the previous subsection,
the simulation domain must be extended on both sides
 through the addition of
three ghost nodes. Let the pair of indices $ij$
( $1 \le i \le 2Q_x$, $1 \le j \le Q_y$) label the momentum
vector corresponding to each discrete population. 

For the bounce-back condition at $x = 0$, the following 
procedure is performed to define the particle populations in the ghost nodes.
Let $F_{1;ij}$, $F_{2;ij}$ and $F_{3;ij}$ be
the population of particles of momentum $\vp_{ij} = (p_{x,i},\,p_{y,j})$, located
in the first three nodes of the simulation domain
near the channel centerline. These
nodes are counted in the positive (right) direction of the $x$ axis.
The first three ghost nodes located at the left of the channel centerline
and counted in the negative direction of the $x$ axis,
have the populations $F_{0;ij}$, $F_{-1;ij}$ and $F_{-2;ij}$, respectively. 
To implement the bounce-back condition, these 
ghost populations are related to the populations in the simulation domain
according  to
\begin{equation}
 F_{0;ij} = F_{1; \widetilde{\imath} \widetilde{\jmath}}, \quad 
 F_{-1;ij} = F_{2; \widetilde{\imath} \widetilde{\jmath}}, \quad 
 F_{-2;ij} = F_{3; \widetilde{\imath} \widetilde{\jmath}},
\end{equation}
where the indices $\widetilde{\imath}$ ($\widetilde{\jmath}$) refer to the
components 
$p_{x,\widetilde{\imath}}$ ($p_{y,\widetilde{\jmath}}$) defined through:
\begin{equation}
 p_{x,\widetilde{\imath}} = -p_{x,i}, \qquad 
 p_{y,\widetilde{\jmath}} = -p_{y,j}.
 \label{eq:widetildei}
\end{equation}

Let $S$ denote the last (rightmost) node located in the flow domain.
The first three ghost nodes outside the right boundary will be denoted
$S+1$,  $S+2$,  $S+3$.  
On the right boundary, the diffuse reflection concept should be imposed.
According to this concept, the flux of particles coming from the ghost
nodes is Maxwellian and equals $\phi^{\rm (eq)}_{w;ij} p_{x,i} / m$,
where $\phi^{\rm (eq)}_{w;ij}$ is defined by Eq.~(\ref{eq:phieq}). 
In the frame of the WENO scheme,
this can be exactly achieved when \cite{busuioc17,busuioc18}:
\begin{equation}
 F_{S+1; ij} = F_{S+2; ij} = F_{S+3; ij} = F^{\rm (eq)}_{w;ij}, \qquad 
 {\mathrm{for}}\,\,\, p_{x,i} < 0,
\end{equation}
where $F^{\rm (eq)}_{w;ij} = \phi^{\rm (eq)}_{w;ij}$ when 
$F_{s; ij}$ refers to $\phi_{s; ij}$, while 
$F^{\rm (eq)}_{w;ij} = \chi^{\rm (eq)}_{w;ij} = T_w \phi^{\rm (eq)}_{w;ij}$
in the case when $F_{s;ij}$ refers to $\chi_{s; ij}$.
Since Eqs.~(\ref{eq:rk3}) and (\ref{eq:flux}) cannot be
used in the nodes $s\in\{S + 1, S+2\}$ when $p_{x,i} > 0$, the 
corresponding functions $F_{s;ij}$, which describe the particles travelling
rightwards, are extrapolated at every time step by a quadratic
procedure:
\begin{multline}
 F_{s; ij} = \frac{(x_s - x_{s-2})(x_s - x_{s-3})}{(x_{s-1} - x_{s-2})(x_{s-1}-x_{s-3})} F_{s-1;ij} \\
 + \frac{(x_s - x_{s-1})(x_s - x_{s-3})}{(x_{s-2} - x_{s-1})(x_{s-2}-x_{s-3})} F_{s-2; ij} \\
 + \frac{(x_s - x_{s-1})(x_s - x_{s-2})}{(x_{s-3} - x_{s-1})(x_{s-3} - x_{s-2})} F_{s-3; ij}.
\end{multline}
The wall density $n_w$, which is required in order to construct 
$F^{\rm (eq)}_{w;ij} \in \{\phi^{\rm (eq)}_{w;ij}, \chi^{\rm (eq)}_{w;ij}\}$,
is thereafter obtained by imposing mass conservation on the right wall:
\begin{equation}
 \sum_{i,j} \Phi_{S + \frac{1}{2}; ij} = 0 \Rightarrow 
 n_w = - \frac{\displaystyle \sum_{i,j, p_{x,i} > 0} \Phi_{S+\frac{1}{2};ij}}
 {\displaystyle \sum_{i,j, p_{x,i} < 0} \frac{\phi^{\rm (eq)}_{w;ij}}{n_w} \frac{p_{x,i}}{m}},
\end{equation}
where $\Phi_{S+1/2;ij}$ is the flux \eqref{eq:weno5_flux} corresponding to the 
reduced distribution $\phi_{ij}$ through the interface between the last 
fluid cell and the first ghost node.

\bibliography{lb}

\end{document}